\documentclass[prd,preprintnumbers,floatfix,
nofootinbib,superscriptaddress,twocolumn]{revtex4-1}
\usepackage{mathtools}
\usepackage{amsfonts} % AMS
\usepackage{amssymb} % AMS
\usepackage{amsmath} % AMS
\usepackage{graphicx} % Include figure files
\usepackage{subfigure} % Include figure files
\usepackage{array} % array
\usepackage{dcolumn} % Align table columns on decimal point
\usepackage{bm} % bold math
\usepackage{latexsym} % latex symbols
\usepackage{longtable} % long tables
\usepackage{hyperref} % hypertext links 
\usepackage{verbatim}
\usepackage{epsfig}
\usepackage{color}
\newcommand{\mh}[0]{\bf \color{red}}

\def\tilde{\widetilde}

\newcommand{\thr}[0]{\textrm{thr}}
\newcommand{\disc}[0]{\textrm{disc}}
\newcommand{\conn}[0]{\textrm{conn}}
\newcommand{\CZ}[2]{C_{#1,\thr}^{(#2)}(0)}
\newcommand{\CT}[2]{\partial_\tau C^{(#2)}_{#1,\thr}(0)}
\newcommand{\CZZ}[1]{C_{#1,\thr}(0)}
\newcommand{\CTT}[1]{\partial_\tau C_{#1,\thr}(0)}

\begin{document}

\title{Perturbative results for two and three particle threshold energies in finite volume}
\author{Maxwell T. Hansen}
\email[Email: ]{hansen@kph.uni-mainz.de}
\affiliation{Institut f\"ur Kernphysik and Helmholz Institute Mainz, Johannes Gutenberg-Universit\"at Mainz,
55099 Mainz, Germany\\
}
\author{Stephen R. Sharpe}
\email[Email: ]{srsharpe@uw.edu}
%
%\homepage[Home page: ]{http://www.phys.washington.edu/users/sharpe/}
%
\affiliation{
 Physics Department, University of Washington, 
 Seattle, WA 98195-1560, USA \\
}
\date{\today}
\begin{abstract}

% SS:new material
We calculate the energy of the state closest to threshold for two and
three identical, spinless particles confined to a cubic spatial volume with periodic boundary conditions and with zero total momentum in the finite-volume frame. The calculation is performed in relativistic quantum field theory with particles coupled via a $\lambda \phi^4$ interaction, and we work through order $\lambda^3$. 
The energy shifts begin at ${\cal O}(1/L^3)$, and we keep 
subleading terms proportional
to $1/L^4$, $1/L^5$ and $1/L^6$.
These terms allow a non-trivial check of
the results obtained from quantization conditions
that hold for arbitrary interactions, namely that of L\"uscher for two particles
and our recently developed formalism for three particles.
We also compare to previously obtained results based on non-relativistic
quantum mechanics. %, and find differences in the $1/L^6$ terms.
\end{abstract}
%
 %
%\pacs{11.80.Gw,12.38.Gc,12.15.-y} ???
%
\maketitle

\section{Introduction}

% SS:new material

We have recently derived a quantization condition that,
subject to some conditions, determines the energy spectrum for
three particles in a cubic box~\cite{Hansen:2014eka,Hansen:2015zga}.
The formalism is fully relativistic, and extends earlier work
in non-relativistic effective field 
theories~\cite{Polejaeva:2012ut,Briceno:2012rv}.
One application of this general result is to determine how the energy of
the state lying closest to the three-particle threshold 
depends on the box size $L$.
We have developed such a threshold expansion through ${\cal O}(1/L^6)$
in Ref.~\cite{HaSh:inprep}. 

This expansion also allows a test of the formalism.
Since the energy shifts are proportional to $1/L^3$,  the kinematics
becomes nonrelativistic for large enough $L$.
Thus we can compare our results to those obtained using 
non-relativistic quantum mechanics in Refs.~\cite{Beane2007,Tan2007}.
This comparison is successful for the three leading terms
(proportional to $1/L^n$ with $n=3, 4, 5$), but turns out not to be useful
for the $1/L^6$ term. 
This is because the three-particle interaction
enters at this order, and there is no {\em a priori} relation 
between a non-relativistic contact interaction
and a relativistic three-particle amplitude at threshold.\footnote{%
One indication that there cannot exist a simple {\em a priori} relation
between the non-relativistic and relativistic three-body interactions
is that both require regularization and are scale and scheme dependent,
with different regularization schemes needed for the two theories.
In fact, in the relativistic context the situation is even more complicated.
This is because, as has been known for a long time
(see, e.g.~Refs.~\cite{Rubin:1966zz,Brayshaw:1969ab,Taylor:1977A,Taylor:1977B},
and our recent discussion in Ref.~\cite{Hansen:2014eka}),
the standard three-particle scattering amplitude diverges at threshold.
To obtain a finite threshold amplitude one must perform subtractions,
e.g.~following the methods introduced in Refs.~\cite{Hansen:2014eka,HaSh:inprep}.
These, however, are not unique---many definitions are possible.}
Thus one can only use the $1/L^6$ terms as a method for determining
this relation, and not as a check on our threshold expansion.

We have thus turned to perturbation theory (PT) as an alternative tool to provide
a test of the $1/L^6$ results from the threshold expansion.
We pick the simplest interacting, perturbatively-renormalizable 
relativistic QFT---scalar $\lambda \phi^4$ theory---and determine the threshold energy shift
through $\mathcal O(\lambda^3)$, keeping terms scaling as $1/L^n$ with $n \leq 6$ in the 
large volume expansion.
Cubic order is sufficient to provide a non-trivial check on the
threshold expansion developed in Ref.~\cite{HaSh:inprep}.
Furthermore, although the $\phi^4$ theory has no bare six-point vertex,
there is an induced three-particle scattering amplitude
starting at ${\cal O}(\lambda^2)$, and the
subtraction methods developed in Refs.~\cite{Hansen:2014eka,HaSh:inprep}
are tested at one-loop order by our calculation. In particular,  as part of our calculation, we have worked out
the subtracted three-particle amplitude at threshold through cubic order,
using the subtraction defined in Ref.~\cite{HaSh:inprep}.

We have carried out the calculation both for two and three particles,
so as to provide further cross-checks. In particular, we can
compare the result of our perturbative threshold expansion for two
particles with that obtained using the relativistic finite-volume 
two-particle formalism developed in Refs.~\cite{Luscher:1986n2,Luscher:1991n1}.
The agreement we find, as described below, gives us confidence in
our methodology.

We can also compare to the results for two particles obtained using non-relativistic quantum
mechanics in Ref.~\cite{Beane2007}. Since the three-particle amplitude does
not enter into this result, the comparison is unambiguous. We find a discrepancy
in terms proportional to $\lambda^2/L^6$,
% and a clue as to its source.
and make a suggestion for its source.

The remainder of this article is organized as follows.
In the following section we give an overview of our  methodology.
Results for the two-particle threshold energy are then worked out in Sec.~\ref{sec:C2}. 
In Sec.~\ref{sec:M3thr} we calculate the subtracted, three-particle scattering amplitude at threshold. 
In Sec.~\ref{sec:C3} we determine the three-particle threshold energy and express it 
in terms of this subtracted amplitude as well as two-to-two scattering observables.
We make the comparisons to prior results in Sec.~\ref{sec:conc},
and present some conclusions.

Three appendices contain technical details. 
Appendix~\ref{app:sums} derives two identities for finite-volume momentum sums.
Appendix~\ref{app:effectiverange} 
works out the s-wave scattering length and effective range in $\lambda \phi^4$ theory.
Finally, appendix~\ref{app:Luscher} develops the threshold expansion
for two particles that follows from L\"uscher's quantization condition
out to ${\cal O}(1/L^6)$.

\section{Overview of calculation and methodology}
\label{sec:methods}

We consider the relativistic QFT described by the Euclidean Lagrangian density 
\begin{multline}
\label{eq:action}
{\cal L} =  \frac12 \partial_\mu\phi\partial_\mu \phi
+ \frac{m^2}2 \phi^2 + \frac{\lambda_0}{4!} \phi^4 \\
+ \frac{\delta Z}2   \partial_\mu\phi\partial_\mu \phi + \frac{\delta Z_m}2  m^2 \phi^2
\,,
\end{multline}
with $\phi$ a scalar field. Although this theory contains no six-particle local interaction,
a $3\to3$ scattering amplitude ${\cal M}_3$ is induced, as discussed below.
This theory has a $\mathbb Z_2$ symmetry, under which $\phi \to -\phi$,
that forbids amplitudes involving an odd number of fields.
This symmetry is also assumed in the general three-body formalism
of Refs.~\cite{Hansen:2014eka,Hansen:2015zga}. 
We have included counterterms for wavefunction and mass renormalization.
These are tuned so that $m$ is the physical, infinite-volume mass and that the residue 
of the infinite-volume propagator at the mass pole is unity.
Note that we do not include an explicit counterterm for the coupling $\lambda_0$, 
preferring to work initially with the bare coupling and later describe its renormalization.

To determine the energies of states close to threshold, we
calculate even and odd particle-number correlators in finite volume.
Choosing convenient overall factors, these are defined by
\begin{align}
\label{eq:C2def}
C_2(\tau) &=\frac{(2m)^2}{2 L^6} e^{2m\tau}
\left\langle \tilde \phi_{\vec 0}(\tau)^2
\tilde \phi_{\vec 0}(0)^2 \right\rangle \,,
\\
\label{eq:C3def}
C_3(\tau) &=\frac{(2m)^3}{6 L^9} e^{3m\tau}
\left\langle \tilde \phi_{\vec 0}(\tau)^3
\tilde \phi_{\vec 0}(0)^3 \right\rangle \,,
\end{align}
where
\begin{equation}
\tilde \phi_{\vec p}(\tau) = \int_L d^3 x \; e^{-i \vec p \cdot \vec x} \phi(\vec x, \tau)\,.
\end{equation}
The subscript ``$L$'' indicates that the spatial
volume is restricted to a cube of side $L$, with periodic boundary
conditions applied to $\phi$.
Thus the allowed momenta are $\vec p=2\pi \vec n/L$,
with $\vec n$ a vector of integers.
We work in Euclidean space, with $\tau$ the Euclidean time,
which is taken to have infinite range.
We choose to place all fields at zero spatial momentum since the
threshold state in the absence of interactions consists of 
particles at rest. Thus our creation and annihilation operators
will have large (${\cal O}(1)$ in PT)
overlap with the actual threshold state even in the presence
of interactions. This is a convenience, but is not strictly necessary
since all we need is for our operators to have some overlap with the threshold
state.

The general form of these two correlators is known in terms of the
eigenstates of the Hamiltonian of the theory. Assuming, as we do henceforth,
that $\tau > 0$, we have
\begin{align}
\label{eq:C2exp}
C_2(\tau) &= \sum_{n\in \text{even}} Z_{2,n} e^{-\Delta E_{2,n}\tau}\,,
\\
\label{eq:C3exp}
C_3(\tau) &= \sum_{n\in \text{odd}} Z_{3,n} e^{-\Delta E_{3,n}\tau}\,,
\end{align}
where
\begin{equation}
\Delta E_{j,n} = E_{j,n}-jm\,.
\label{eq:DeltaEjdef}
\end{equation}
Due to the $\mathbb Z_2$ symmetry, it is possible to separate states 
with even- and odd-particle quantum numbers into $C_2$ and $C_3$ respectively.
This implies that $C_2$ contains a 
contribution from the vacuum state, which [given the inclusion of
the $\exp(2m\tau)$ in its definition, Eq.~(\ref{eq:C2def})]
leads to a growing exponential of $\tau$.
Similarly, $C_3$ contains an exponentially growing
contribution from a single-particle state.
Such growing exponentials might be problematic in a numerical simulation,
but can be readily identified in an analytic calculation.

Our aim is to pick out from the infinite sum of exponential contributions,
those corresponding to the states nearest threshold. 
We know that there is only one such state for each correlation function,
since there is only one such state in the free theory (with all particles at rest)
and we are making an infinitesimal perturbation.
For these threshold states,
the quantities  $\Delta E_{j,n}$
vanish as $L\to\infty$ as a sum of powers of $1/L$, 
up to possible logarithmic corrections---which in fact do not arise at the order we 
work---and exponentially suppressed terms. The latter, which behave as $e^{-mL}$,
we neglect throughout. 
Such corrections are also dropped in L\"uscher's general two-particle quantization 
condition and in our general three-particle formalism.
As mentioned above, in this work we expand the threshold energy shift, $\Delta E_{j,n}$, 
in both $\lambda_0$ and $1/L$, working through $\mathcal O(\lambda_0^3)$ and $\mathcal O(1/L^6)$. 

As is well known (see, e.g., 
Refs.~\cite{Luscher:1986n2,Luscher:1991n1,Beane2007,Tan2007}),
the leading contribution to the threshold energy shift is proportional to $1/L^3$. 
As we now explain, this implies that we can unambiguously pick out from the $C_j(\tau)$
the contribution from the near-threshold state.
Consider first the behavior of $\Delta E_{j,n}$ from excited states.
The lightest such state adds a minimal unit of relative momentum between two
of the particles, so that
$\Delta E_{2,n} = 2 \omega_p - 2 m + \mathcal O(\lambda_0)$,
where $\omega_p \equiv \sqrt{\vec p^{\,2} + m^2}$ with $|\vec p|=2 \pi/L$. 
For $L\gg 1/m$, the energy shift can be expanded as
$\Delta E_{2,n} \propto \vec p^2 \propto 1/L^2$, and so we see that the
excited states are parametrically separated from the near-threshold state, whose
energy shift scales as $1/L^3$ as noted above. Thus there can be no avoided level crossings,
and, in an analytic calculation, we can unambiguously identify the excited state contributions.
We stress that it is crucial to discard these exponentials before expanding in $1/L$,
so that the contributions do not become confused with the ground-state energy shift.
Excited states involving more particles 
(e.g. five-particle states in $C_3$)
are even more obviously separated, since then $\Delta E_{j,n} \approx 2 m$,
which does not vanish as $L\to \infty$. 
Similarly, far-subthreshold states, which contain less particles (e.g  single particle
states in $C_3$), have $\Delta E_{j,n}\approx - 2m$, and the exponentials can also
be easily separated.

In light of the previous discussion, the method we use is as follows.
We calculate $C_2$ and $C_3$ order by order in PT, and remove by hand
the contributions from exponentially growing far-subthreshold states and
from exponentially falling excited states. 
The resulting subtracted correlators we call $C_{j,\thr}$. 
We know that these have the form
\begin{equation}
C_{j,\thr} = Z_{j,\thr} e^{-\Delta E_{j,\thr} \tau}
\,.
\label{eq:Cthr}
\end{equation}
Thus if we expand in powers of $\tau$, and keep only the constant
and linear terms,
\begin{equation}
C_{j,\thr}(\tau) = \CZZ j + \tau \big [ \CTT j \big ] + {\cal O}(\tau^2)\,,
\label{eq:Cjexpand}
\end{equation}
%\begin{equation}
%C_{j,\thr}(\tau) = C_{j,\thr}^{(0)} + \tau C_{j,\thr}^{(1)} + {\cal O}(\tau^2)\,,
%\label{eq:Cjexpand}
%\end{equation}
then the threshold energy shift is given by
\begin{equation}
\Delta E_{j,\thr} = - \frac{\CTT j}{\CZZ j}\,.
\label{eq:DeltaEmethod}
\end{equation}
The advantage of this method is that it allows us to keep track of powers
of $\lambda_0$ in a straightforward manner.
An alternative approach would be to identify the infinite set of
perturbative diagrams leading to the exponential
behavior in Eq.~(\ref{eq:Cthr}), with its associated renormalization
factor $Z_{j,\thr}$. This requires working to all orders in $\lambda_0$
in a subset of diagrams. We have used this alternate
method as a check on our results, though we present no details here.

\begin{figure}[tbh]
\begin{center}
\includegraphics[scale=0.46]{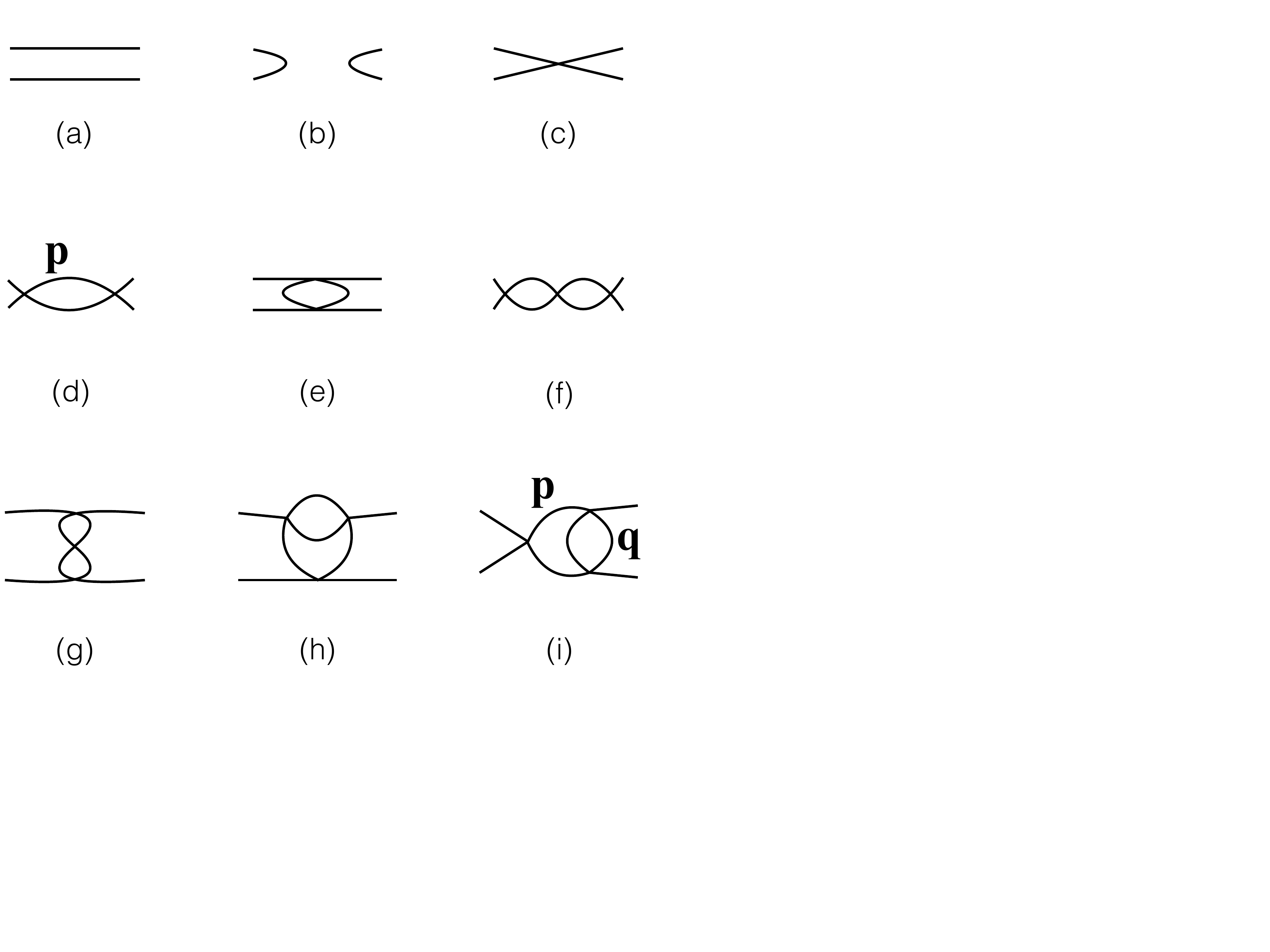}
\vskip -1.4truein
\caption{Feynman diagrams contributing to the even particle-number correlation function. 
External particles have zero three-momentum. 
Diagram (i) also has a horizontally flipped partner, not shown.
Examples of labels used in the text for loop momenta are shown.}
\label{fig:C2}
\end{center}
\end{figure}

\begin{figure*}[tbh]
\begin{center}
\includegraphics[scale=0.5]{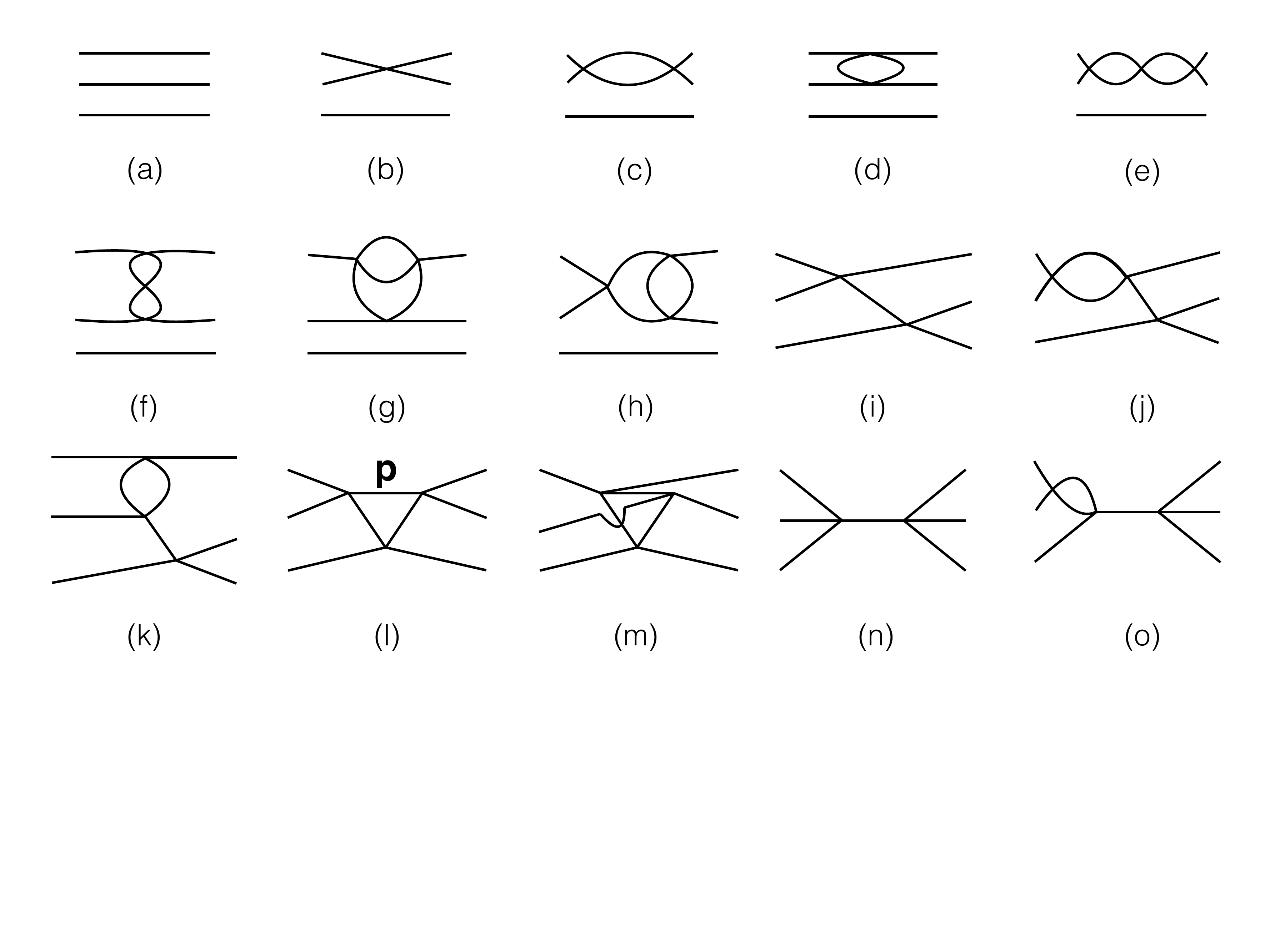}
\vskip -1.5truein
\caption{Feynman diagrams contributing to the odd particle-number 
correlation function $C_3$. External particles have vanishing spatial momentum. 
Not shown are diagrams obtained from (h), (j), (k) and (o)
by applying the loop correction to the other vertex.
An example of loop-momentum labeling is shown.
Figs.~(i)-(o) are also diagrams for the connected part of the 
three-particle scattering amplitude. We reference these diagrams, reinterpreted as infinite-volume scattering contributions, in our calculation of Sec.~\ref{sec:M3thr}. For that calculation the external particles are not assumed to be at rest.}
\label{fig:C3}
\end{center}
\end{figure*}

The diagrams that we need to calculate to obtain $\Delta E_{j,\thr}$ up to
third order in $\lambda_0$ are shown in Figs.~\ref{fig:C2} and
\ref{fig:C3} for $C_2$ and $C_3$, respectively. 
The free propagators are 
\begin{equation}
\langle \tilde \phi_{\vec p}(\tau_1) \phi(\vec 0,\tau_2) \rangle_{\lambda_0=0}
= \frac{1}{2 \omega_{ p}} e^{- \omega_{ p} |\tau_1-\tau_2|} \,,
\label{eq:prop}
\end{equation}
where, again, $\omega_p=\sqrt{\vec p^{\,2}+m^2}$ is the energy
of a free particle with a finite-volume momentum $\vec p$.
For each loop there will be a sum over spatial momentum restricted
to the allowed finite-volume values.
In addition there will be an integral over the Euclidean time of each
intermediate vertex. For given values of each of the spatial momenta,
this is simply an integral over exponentials, with the integrand depending
on the time ordering due to the absolute values in the propagators, 
Eq.~(\ref{eq:prop}).
Thus the integrals are trivial, but keeping track of all the time orderings
is less so. We have used two independent Mathematica codes
to ensure that all terms are included.
After the vertex integrals, one can read off which terms are exponentially
falling or growing, drop these by hand, and thus obtain $C_{j,\thr}(\tau)$.
Expanding in powers of $\tau$ leads to the results for $\CZZ j$
and $\CTT j$ [see Eq.~(\ref{eq:Cjexpand})]
and thus to $\Delta E_{j,\thr}$ [using Eq.~(\ref{eq:DeltaEmethod})].

The final stage of the calculation is, for each loop, to sum 
over the finite-volume momenta.
This is done by converting sums to integrals, which can be absorbed
into infinite-volume loop contributions, 
plus a finite-volume residue leading to the desired power-law terms. 
The methods for converting sums to integrals are variants and extensions of
those used in deriving the general finite-volume quantization 
conditions~\cite{Luscher:1986n2,Luscher:1991n1,Hansen:2014eka}
and their threshold expansions~\cite{HaSh:inprep}.
The results we need are collected in Appendix~\ref{app:sums}.

As a simple illustration of these methods, consider 
the diagrams of Figs.~\ref{fig:C2}(a) and \ref{fig:C3}(a).
By construction, both diagrams lead to $C_j^{({a})}(\tau) = 1$. 
In particular, the contraction factors cancel the $1/2$ and $1/6$ in Eqs.~(\ref{eq:C2def})
and (\ref{eq:C3def}), respectively.
There are thus no exponentially growing or falling terms to remove by hand, 
and $C_{j,\thr}(\tau)=C_j(\tau)=1$ at this order.

Since only the constant term in $C_{j,\thr}$ is non-vanishing at leading order,
the perturbative expansions of the quantities of interest can be written:
\begin{align}
\CZZ j &= 1 + \sum_{n=1}^\infty  \lambda_0^n \CZ j n \,,
\label{eq:Cjthr0exp}
\\
\CTT j &= \sum_{n=1}^\infty  \lambda_0^n \big  [\CT j n \big ] \,,
\label{eq:Cjthr1exp}
\\
\Delta E_{j,\thr} &= \sum_{n=1}^\infty  \lambda_0^n \Delta E_{j,\thr}^{(n)} \,.
\label{eq:DeltaEexp}
\end{align}
%\begin{align}
%C_{j,\thr}^{(0)} &= 1 + \sum_{n=1}^\infty  \lambda_0^n \CZ j n \,,
%\label{eq:Cjthr0exp}
%\\
%C_{j,\thr}^{(1)} &= \sum_{n=1}^\infty  \lambda_0^n \big  [\CT j n \big ] \,,
%\label{eq:Cjthr1exp}
%\\
%\Delta E_{j,\thr} &= \sum_{n=1}^\infty  \lambda_0^n \Delta E_{j,\thr}^{(n)} \,.
%\label{eq:DeltaEexp}
%\end{align}
Inserting these expansions into Eq.~(\ref{eq:DeltaEmethod}), we find
\begin{align}
\Delta E_{j,\thr}^{(1)} &= - \CT j 1 \,,
\label{eq:DeltaE1def}
\\
\Delta E_{j,\thr}^{(2)} &= - \CT j 2 + \CZ j 1 \big [\CT j 1 \big ]\,,
\label{eq:DeltaE2def}
\\
\Delta E_{j,\thr}^{(3)} &= - \CT j 3 + \CZ j 1 \big [\CT j 2 \big ]
\nonumber\\
&  \hspace{5pt} + \left( \CZ j 2 - {\CZ j 1}^2  \right ) \big [\CT j 1 \big ]
\,.
\label{eq:DeltaE3def}
\end{align}
%\begin{align}
%\Delta E_{j,\thr}^{(1)} &= - C_{j,\thr}^{(1,1)}\,,
%\label{eq:DeltaE1def}
%\\
%\Delta E_{j,\thr}^{(2)} &= - C_{j,\thr}^{(1,2)} + C_{j,\thr}^{(0,1)} C_{j,\thr}^{(1,1)}\,,
%\label{eq:DeltaE2def}
%\\
%\Delta E_{j,\thr}^{(3)} &= - C_{j,\thr}^{(1,3)} + C_{j,\thr}^{(0,1)} C_{j,\thr}^{(1,2)}
%\nonumber\\
%&\quad + C_{j,\thr}^{(0,2)} C_{j,\thr}^{(1,1)} 
%- \left(C_{j,\thr}^{(0,1)}\right)^2 C_{j,\thr}^{(1,1)}\,.
%\label{eq:DeltaE3def}
%\end{align}
Thus a third-order calculation of $\Delta E_{j,\thr}$ requires a
third order result for $\CTT j$ but only a second-order result for
$\CZZ j$.

Another simple example of our methods is provided by the
disconnected diagram of Fig.~\ref{fig:C2}(b).
This diagram clearly has no dependence on $\tau$, so when we use
Eq.~(\ref{eq:C2def}) we find that $C_2(\tau) \propto e^{2 m \tau}$.
Dropping this exponentially growing term leads to $C_{2,\thr}=0$, so this
diagram makes no contribution to the threshold energy.

The final example we consider here is the
connected diagram Fig.~\ref{fig:C2}(c), which involves a single vertex.
All four propagators have vanishing spatial momenta, so we only need to
do the integral over the vertex position, $\tau_1$. The resulting
contribution to the correlator is
\begin{align}
C_2^{(c)}(\tau) &= - \frac12\frac{\lambda_0}{(2m)^2} \frac{1}{L^3}
\Big\{ \int_0^\tau d\tau_1 \;1 
\nonumber\\
&\ \ + \int_{-\infty}^0 d\tau_1\; e^{4 m \tau_1}
+ \int_\tau^{\infty} d\tau_1\; e^{-4 m (\tau_1-\tau)} \Big\}
\\
&=- \frac{\lambda_0}{8m^2 L^3} \left(\frac1{2m} + \tau\right)
\,.
\end{align}
Here there are no exponentially growing or falling terms to remove,
and we find the first nontrivial result for the threshold energy shift
\begin{align}
\CT 2 1 &= - \frac{1}{8 m^2 L^3} = -\Delta E_{2,\thr}^{(1)}
\label{eq:DeltaE21}
\,,
\end{align}
as well as an ${\cal O}(\lambda_0)$ contribution to the constant term
\begin{align}
\CZ 2 1 &= - \frac{1}{16 m^3 L^3} \,.
\label{eq:C201}
\end{align}

\begin{figure}[tbh]
\begin{center}
\includegraphics[scale=0.45]{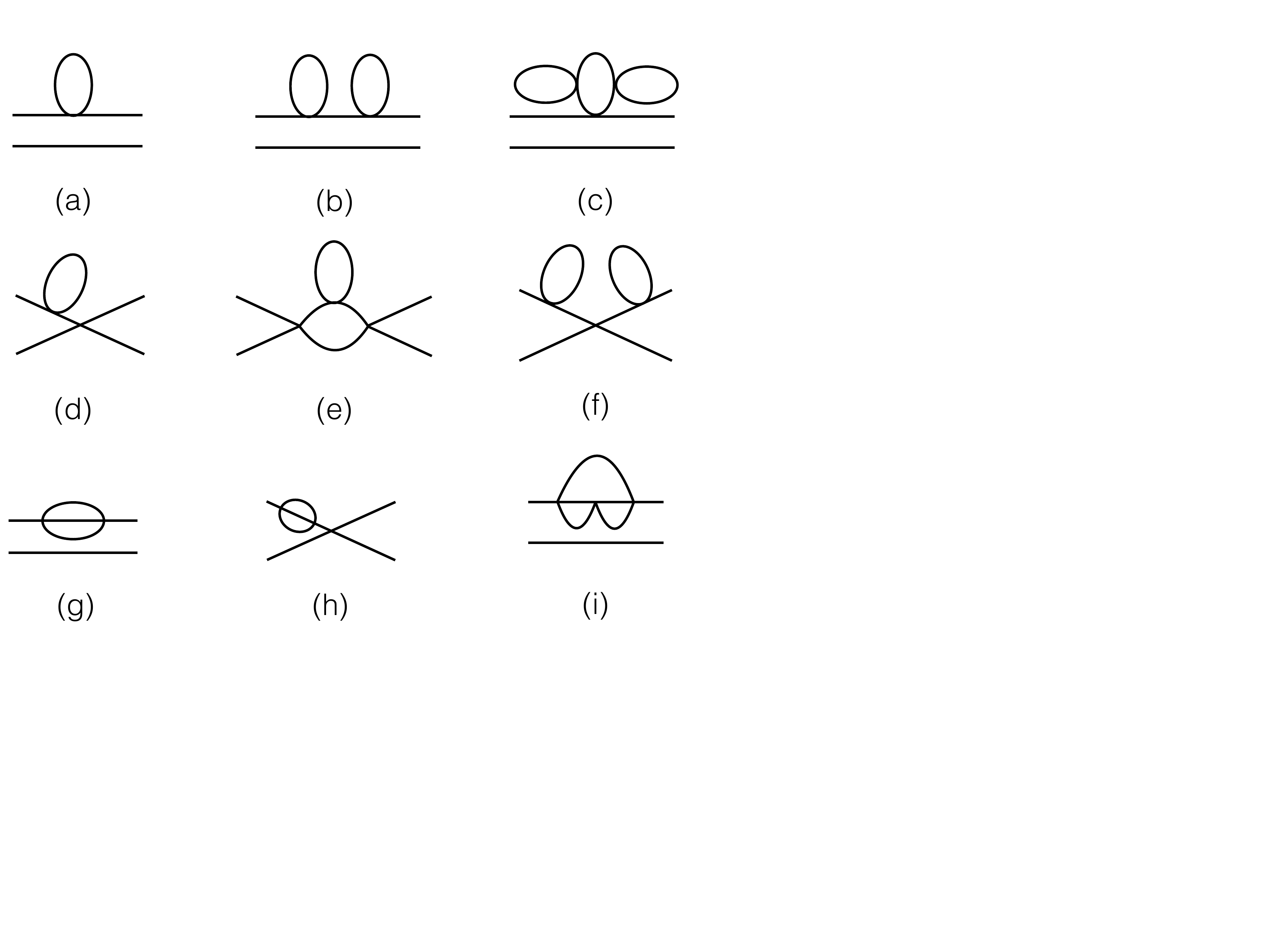}
\vskip -1.5truein

\caption{Examples of Feynman diagrams leading to mass and wavefunction renormalization
in the correlator $C_2$ at the order we work. 
Similar diagrams involving the counterterms $\delta Z$ and $\delta Z_m$ are not shown.}
\label{fig:renorm2}
\end{center}
\end{figure}

\bigskip

\begin{figure}[tb]
\begin{center}
\vspace{10pt}
\includegraphics[scale=0.45]{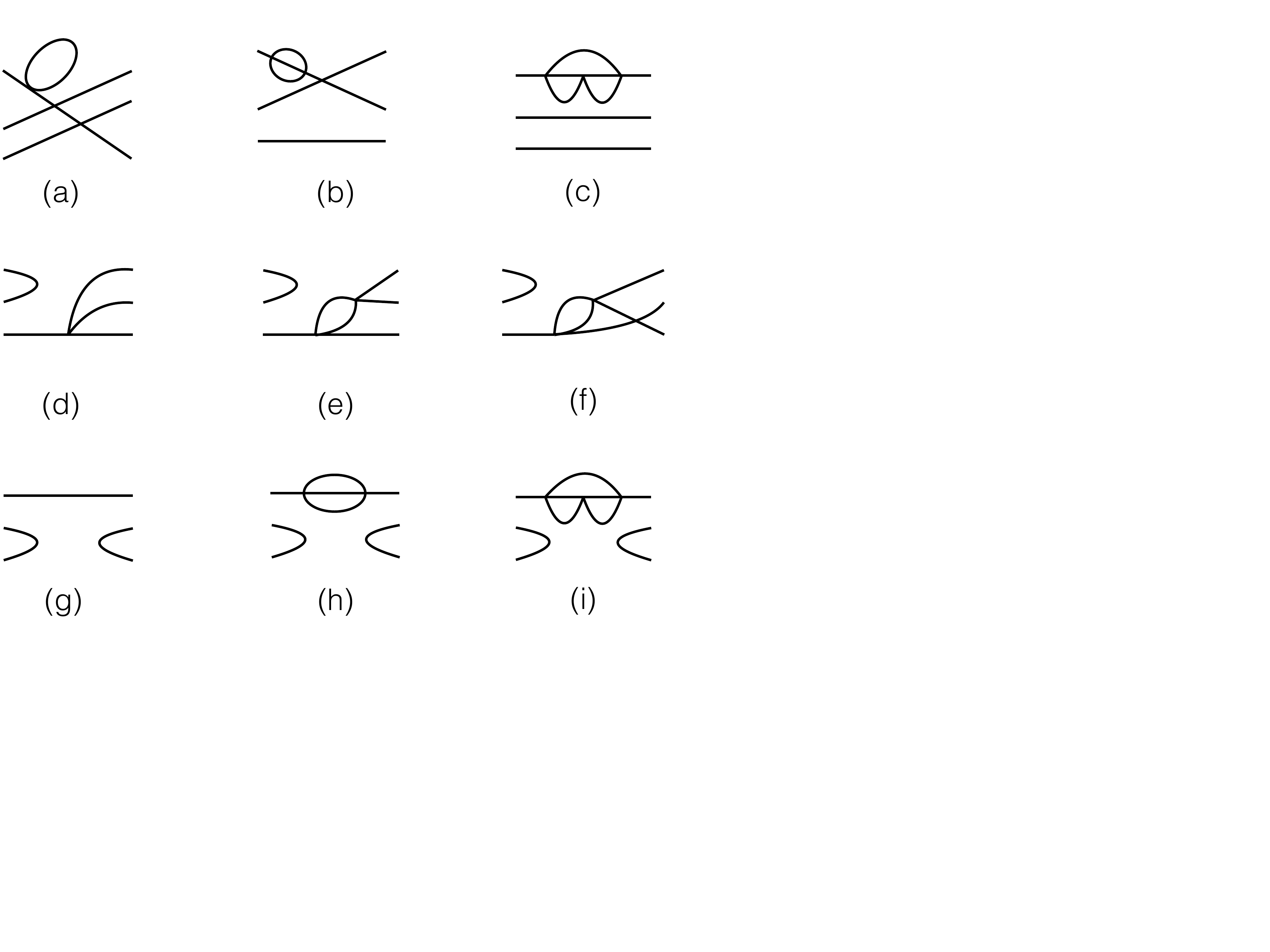}
\vskip -1.5truein
\caption{Examples of Feynman diagrams whose contributions to $C_3$ lead only to renormalizations of the constant
$Z_{3,\thr}$ but not to changes in the threshold energy shift $\Delta E_{3,\thr}$. Diagrams similar to (a)-(c) but 
involving counterterms are not shown.}
\label{fig:renorm3}
\end{center}
\end{figure}

We now describe three classes of diagram that we do not need to calculate explicitly,
although they contribute to the $C_j(\tau)$ at the order we work.
This is because they either do not contribute to $C_{j,\thr}(\tau)$, or
they lead only to changes in the overall normalization of the correlators,
$Z_{j,\thr}$, but not to the energy shifts $\Delta E_{j,\thr}$.
Examples of these diagrams are shown in Fig.~\ref{fig:renorm2} for $C_2$ and
Fig.~\ref{fig:renorm3} for $C_3$. 

The first class involves self-energy and counterterm insertions on the diagrams
 described above (i.e., those in Figs.~\ref{fig:C2} and \ref{fig:C3}).
All diagrams in Fig.~\ref{fig:renorm2} and those of Fig.~\ref{fig:renorm3} (a)-(c) are examples of this class.
We first note that contributions involving tadpole diagrams, such as those in Figs.~\ref{fig:renorm2}(a)-(f) and
Fig.~\ref{fig:renorm3}(a), are cancelled  identically  by the $\delta Z_m$ counterterm.
This is because the loops are independent of the external momenta and thus lead only to mass renormalization.
The only subtlety is that $\delta Z_m$ actually cancels the infinite-volume version of the tadpole diagram,
in which the loop is integrated rather than summed. However, since the difference between the sum and the integral is exponentially suppressed, scaling like $e^{-mL}$, no contribution remains in our $1/L$ expansion.\footnote{%
The difference between loop sums and integrals is exponentially suppressed as long as there are no
cuts through the loops in which, for threshold kinematics, all particles can go on shell.
Several examples where this is not the case are described in subsequent sections.}

The other renormalization diagrams that contribute at the order we work are those that do not involve
tadpoles. These 
are those shown in Figs.~\ref{fig:renorm2}(g)-(i) and Figs.~\ref{fig:renorm3}(b) and (c),
together with related diagrams in which the renormalizations appear on different external lines,
and diagrams containing the corresponding counterterms.
Note that  renormalizations appear only on external propagators at this order.
For all these diagrams the loop sums can be replaced
by integrals---there are no power-law finite-volume residues. Thus, if the propagators were
evaluated on mass shell, the contributions of these diagrams would exactly cancel with those
containing the appropriate mass and wavefunction counterterms (given our convention that the residue at
the mass pole is unity). In fact, the external propagators, while evaluated at vanishing spatial momenta,
are not Fourier transformed in time. Thus they contain excited state contributions. 
However, once these are removed, following our general prescription described above,
we expect the general argument to hold and the contributions of these diagrams and those with the
corresponding counterterms to cancel exactly. We have checked that this is the case by explicitly
calculating these diagrams.

The second class of diagrams are those exemplified by Figs.~\ref{fig:renorm3}(d), (e) and (f),
as well as those obtained by horizontal reflection.
The characteristic feature of this class is that there is a disconnected propagator joining two of the
external fields at either the initial or final time, multiplied by a ``one-to-three'' correlator.
This factorization is maintained as higher order corrections are included.
The disconnected propagator has no $\tau$ dependence, and so provides only an overall factor.
Thus, applying our methodology to this class of diagrams amounts to studying the three-particle
threshold energy using the one-to-three correlator. As noted above, this is a legitimate approach,
since one can use any interpolating fields with the correct quantum numbers. 
In particular, this class of diagrams alone must give the same result for $\Delta E_{3,\thr}$
as that obtained from the full $C_3$. Thus we can drop these one-to-three diagrams without
changing the result for the energy shift.
We have checked this argument by explicitly calculating all the diagrams in this class up to
order $\lambda_0^3$. Note that, since one vertex is needed to convert the initial single particle
into three, only a second-order result for $\Delta E_{3,\thr}$ is obtained.

The third and final class of diagrams are those  exemplified by Figs.~\ref{fig:renorm3}(g), (h) and (i).
Here one is effectively calculating the threshold energy shift using the ``one-to-one'' correlator,
with the disconnected propagators at each end only changing the overall factor.
Once again, this class of diagrams alone, analyzed using our method, must yield the correct
result for $\Delta E_{3,\thr}$, and so can be dropped. 

\section{Two particle energy shift}
\label{sec:C2}

In this section we calculate the threshold energy shift for two particles.
We work to ${\cal O}(\lambda_0^3)$ in PT
and keep terms up to $\mathcal O(1/L^6)$ in the volume expansion.
We have already obtained the contribution linear in $\lambda_0$,
Eq.~(\ref{eq:DeltaE21}), so we start here with the quadratic term,
which arises from Figs.~\ref{fig:C2}(d) and (e).

For Fig.~\ref{fig:C2}(d), the form of the result depends on whether the
loop momentum, $\vec p$, vanishes or is non-vanishing.
This is because one of the terms which enters the energy shift in the $\vec p=0$ case becomes an excited state exponential, to be discarded, in the case of non-vanishing loop momenta. 
The zero-momentum contribution is
\begin{equation}
C_2(\tau) \supset  \left(\frac{\lambda_0}{8 m^2 L^3}\right)^2
\left[\frac3{8m^2} + \frac{3\tau}{4m} + \frac{\tau^2}2\right]\,.
\label{eq:C2d0}
\end{equation}
The $\tau^2$ term is the second term in the expansion of 
$\exp(-\Delta E_{2,\thr}^{(1)}\tau)$,
with $\Delta E_{2,\thr}^{(1)}$ given in Eq.~(\ref{eq:DeltaE21}),
and does not concern us here.
\begin{comment}
The term linear in $\tau$ contributes to $C_{2,\thr}^{(1,2)}$ as
\begin{equation}
C_{2,\thr}^{(1,2)}\supset  \frac{3 m}{(16 m^3 L^3)^2}\,.
\end{equation}
We do not need, however, to keep the constant term in Eq.~(\ref{eq:C2d0}). 
This gives a contribution to $C_{2,\thr}^{(0,2)}$ proportional to $1/L^6$. As can be seen from 
Eq.~(\ref{eq:DeltaE3def}), this must be multiplied by $C_{j,\thr}^{(1,1)}\sim 1/L^3$ to
contribute to the energy shift, and thus gives $\Delta E_{j,\thr}^{(3)}\sim 1/L^9$.
This is of higher order than the $1/L^6$ terms we are aiming for.
\end{comment}
For $\vec p\ne 0$ we find the contribution
\begin{multline}
C_2(\tau) \supset \frac{\lambda_0^2}{64 m^2 L^3} \frac1{L^3}\sum_{\vec p\ne 0}^\Lambda 
\frac1{\omega_p^2}
\bigg[ \frac{\omega_p(\omega_p^2-3m^2)}{2 m(\vec p^2)^2}
+ \tau \frac{\omega_p}{\vec p^2}  \\
 +{e^{-2(\omega_p-m)\tau}}  \frac{m^2}{(\vec p^{2})^2}  \bigg ]
\,.
\label{eq:C2d1}
\end{multline}
The superscript $\Lambda$ indicates that ultraviolet (UV) regularization is required,
although the choice of regulator is unimportant.
The last term in the summand is from excited states, 
and is dropped in $C_{2,\thr}(\tau)$, as explained in the previous section. 
The remaining two parts of the summand diverge for $\vec p \rightarrow 0$,
and are converted to integrals plus finite-volume residues
using the results in Appendix~\ref{app:sums}.
For example, the term proportional to $\tau$ leads to the following
contribution to $\CT 2 2$:
\begin{align}
\CT 2 2 &\supset \frac1{64m^2L^3} \frac1{L^3}\sum_{\vec p\ne 0}^\Lambda
\frac1{\omega_p \vec p^2}
\\
& \hspace{-50pt} = 
\frac1{64m^2L^3}
\left\{ \int_{\vec p}^\Lambda \left(\frac1{\omega_p \vec p^2}\right)
+\frac{\cal I}{4\pi^2 mL} + \frac1{2 m^3 L^3}\right\}
\,,
\label{eq:C212aux}
\end{align}
where we have used Eq.~(\ref{eq:SIres}) and introduced the shorthand 
$\int_{\vec p} \equiv \int d^3 p/(2 \pi)^3$. 

Turning to Fig.~\ref{fig:C2}(e), we note that $\vec p=0$ does not
need to be treated separately.
After dropping excited-state contributions, we obtain
\begin{equation}
C_{2,\thr}(\tau) \supset
\frac{\lambda_0^2}{32 m^2 L^3} \frac1{L^3}\! \sum_{\vec p}^\Lambda
\left(
\frac{\omega_p^2+\omega_p m - m^2}{2m\omega_p^4(\omega_p+m)}
%\frac{1}{2m\omega_p^3} - \frac{m}{2 \omega_p^4(\omega_p+m)}
+ \frac{\tau}{\omega_p^3}\right)
\,.
\label{eq:C2e}
\end{equation}
Here the sum can be replaced by an integral, since the summand is 
nonsingular and we are dropping exponentially suppressed volume dependence.

Combining the results from Eqs.~(\ref{eq:C2d0}), (\ref{eq:C2d1}) and
(\ref{eq:C2e}),
and evaluating the sums using Eqs.~(\ref{eq:SIres}) and (\ref{eq:SJres}), 
we obtain
\begin{align}
\hspace{-5pt} \CT 2 2 & = 
%\frac1{64m^2 L^3}\left(8 A_2
%%\int_{\vec p}^\Lambda \left(\frac1{\omega_p \vec p^2} + \frac2{\omega_p^3}\right)
%+ \frac{\cal I}{4 \pi^2 mL} + \frac5{4m^3 L^3} \right)
\frac{A_2}{8 m^2 L^3}
+ \frac{\cal I}{256 \pi^2 m^3 L^4} 
+ \frac5{256m^5 L^6} 
\,,\label{eq:C212}
\\
\begin{split}
\CZ 2 2 & = 
\!-  \frac{\cal J}{1024 \pi^4 m^2 L^2}
\!+\! \frac{A_2}{16 m^3 L^3} 
\\ & \hspace{80pt}  \!+\! \frac{I^{(0,2)}}{64m^3L^3} 
\!+\! {\cal O}(L^{-4})
\,.\label{eq:C202}
\end{split}
\end{align}
Here $A_2$ is the one-loop integral defined in Appendix B, Eq.~(\ref{eq:A2def}),
while $I^{(0,2)}$ is the finite integral
\begin{align}
I^{(0,2)} &= \int_{\vec p} \frac{m}{\omega_p+m} \left[
\frac1{\omega_p \vec p^2}- \frac{m}{\omega_p^4}\right]
= \frac1{8\pi}
\,.
\end{align}

Two comments are in order. 
First, the $A_2$ terms are exactly those needed to convert
the factor of $\lambda_0$ multiplying the first-order results
in Eqs.~(\ref{eq:DeltaE21}) and (\ref{eq:C201}) into the renormalized $\lambda$.
The latter is defined in Eqs.~(\ref{eq:lambdadef})-(\ref{eq:lambdadef2}), which we reproduce
here for clarity:
\begin{equation}
 \lambda =  32 m \pi a = \lambda_0 - A_2 \lambda_0^2 + A_3 \lambda_0^3+ \mathcal O(\lambda_0^4)\,,
\end{equation}
where $a$ is the two-particle scattering length.
Second, we have truncated $\CZ 2 2$ at ${\cal O}(1/L^3)$,
since higher order terms in $1/L$ lead to contributions to
$\Delta E_{2,\thr}^{(3)}$ of ${\cal O}(1/L^7)$.
This is because $\CZ 2 2$ multiplies $\CT 2 1 \propto 1/L^3$
when it contributes to $\Delta E_{2,\thr}^{(3)}$, as can be seen from
Eq.~(\ref{eq:DeltaE3def}).

Now we move to third order. As is clear from Eq.~(\ref{eq:DeltaE3def}),
we only need to determine the term linear in $\tau$, $\CT 2 3$,
in order to obtain $\Delta E_{2,\thr}^{(3)}$. Furthermore, we can drop
any contributions falling as $1/L^7$ or faster. % in $C_{2,\thr}^{(1,3)}$.

We begin with Fig.~\ref{fig:C2}(f). If both loop momenta vanish,
the diagram is proportional to $1/L^9$ and can be dropped.
If only one loop momentum vanishes, we find (dropping higher order terms)
\begin{align}
\CT 2 3 &\supset \frac{1}{1024 m^5 L^6} \frac1{L^3}\! 
\sum_{\vec p\ne 0}^\Lambda \frac{4m^2-3\vec p^2}{\omega_p (\vec p^2)^2} \,,
\\
\begin{split}
& \hspace{-15pt} =\frac{\mathcal J}{4096 \pi^4 m^4 L^5}  \\
& \hspace{10pt} - \frac{1}{256 m^5 L^6}
\left(I^{(1,3)} + \int_{\vec p}^\Lambda \frac3{4\omega_p \vec p^2}\right) \,,
\label{eq:C2f}
\end{split}
\end{align}
where 
\begin{align}
I^{(1,3)} &= \int_{\vec p} \frac{m}{\omega_p(\omega_p+m) \vec p^2}
= \frac1{2\pi^2}
\,.
\end{align}
The UV divergent integral in Eq.~(\ref{eq:C2f}) 
combines with that in the result from Fig.~\ref{fig:C2}(i)
[given in Eq.~(\ref{eq:C2i}) below] to give a term proportional to $A_2$.
This turns out to be exactly the contribution needed to convert the factor of
$\lambda_0^2$ multiplying the $\tau$ term in Eq.~(\ref{eq:C2d0}) to $\lambda^2$.

If both momenta are non-vanishing, the result factorizes into a product
of loop sums\footnote{%
This factorization occurs because, in order to obtain a contribution proportional
to $\tau$, the times of the vertices must satisfy
$0< \tau_1,\tau_2,\tau_3 < \tau$.  The set-up is then
essentially the same as in the calculation of a threshold scattering amplitude, 
for which we know, from using Feynman diagrams, that the contributions
from the two loops factorize. This is true both in finite and infinite volume.}.
\begin{align}
\CT 2 3 &\supset -\frac1{512 m^2 L^3} 
\left(\frac1{L^3}\sum_{\vec p\ne0} \frac1{\omega_p \vec p^2}\right)^2 \,,
\\
& \hspace{-50pt} = -\frac1{512 m^2 L^3} 
\left(\int_{\vec p} \frac1{\omega_p \vec p^2}
+\frac{\cal I}{4\pi^2 mL} + \frac1{2m^3L^3}\right)^2
\,.
\end{align}
The only product of finite-volume residues that
falls slowly enough to be included is
\begin{equation}
\CT 2 3 \supset - \frac{\mathcal I^2}{8192 \pi^4 m^4 L^5}
\,.
\label{eq:C2fb}
\end{equation}
Terms involving a single finite-volume residue multiplied by the integral
give part of the contribution needed to convert
the factor of $\lambda_0^2$ multiplying the ${\cal I}/L^4$ and
$1/L^6$ terms in Eq.~(\ref{eq:C212aux}) into $\lambda^2$.
Terms involving two integrals contribute to 
two-loop renormalization, generating part of the $A_3 \lambda_0^3$ term 
which converts $\lambda_0$ to $\lambda$ in the $\lambda_0/L^3$ contribution to $C_{2,\thr}$.

The sums in both loops in Fig.~\ref{fig:C2}(g) and (h) can all be converted
into integrals, and these integrals
are exactly those obtained when the same diagrams
are evaluated as contributions to infinite-volume scattering.
It follows that  these diagrams contribute only to renormalization
of lower-order terms.

This leaves Fig.~\ref{fig:C2}(i), and its horizontal reflection.
Here we must treat the cases $\vec p=0$ and $\vec p \neq 0$ separately 
(see the label in the figure), since the separation between ground and 
excited states is different in the two cases. For $\vec p = 0$, the contribution 
to the threshold correlator is 
\begin{align}
\hspace{-15pt} \CT 2 3 &\supset
\frac{1}{512 m^5 L^6} \frac1{L^3}\!\sum_{\vec q}^\Lambda
%\frac{3\omega_q^2+3\omega_qm-2 m^2}{\omega_q^4 (\omega_q+m)}
\left( \frac{2 m^2}{\omega_q^4(\omega_q+m)}
-\frac{3}{\omega_q^3} \right) \,,
\\
&= 
\frac{4-\pi}{2048\pi^2m^5 L^6}
- \frac{1}{512 m^5 L^6} \int_{\vec q}^\Lambda\frac{3}{\omega_q^3}
\,.
\label{eq:C2i}
\end{align}
The UV divergent integral leads to a renormalization of
lower-order terms, as described above in the discussion
following Eq.~(\ref{eq:C2f}). 

For $\vec p\ne 0$, the diagram contributes
\begin{align}
\CT 2 3 &\supset
- \frac{\lambda_0^3}{64 m^2 L^3} \frac{1}{L^6}\! \sum_{\vec p\ne 0,\vec q}^\Lambda
\frac{G(\vec p,\vec q)}{\vec p^2} \,,
\\
G(\vec p,\vec q) &=
\frac{\omega_p + W}{\omega_p \omega_q \omega_{pq} (W^2-m^2)}
\,,\label{eq:Gpq}
\end{align}
where $W=\omega_p\!+\!\omega_q\!+\!\omega_{pq}$ and
$\omega_{pq} = \sqrt{(\vec p\!+\!\vec q)^2+m^2}$.
We can replace the sum over $\vec q$ with an integral
since the summand is regular, leading to
\begin{align}
\CT 2 3 &\supset
- \frac{\lambda_0^3}{64 m^2 L^3} \frac{1}{L^3}\! \sum_{\vec p\ne 0}^\Lambda
\frac{f(\vec p^2)}{\vec p^2}
\\
& \hspace{-50pt} =
- \frac{\lambda_0^3}{64 m^2 L^3} \left(
\int_{\vec p,\vec q} \frac{G(\vec p,\vec q)}{\vec p^2}
+ \frac{\mathcal I f(0)}{4\pi^2L}
- \frac{f'(0)}{L^3}\right)
\,,
\label{eq:C2ib}
\end{align}
where
\begin{equation}
f(\vec p^{\,2}) = \int_q^\Lambda G(\vec p, \vec q)
\,.
\end{equation}
Here we have assumed that the UV regulator maintains rotational
invariance, and used Eq.~(\ref{eq:SIres}).
The first term in Eq.~(\ref{eq:C2ib}) contributes to the
two loop renormalization of $\CT 2 1$,
while the second completes the renormalization of
the $\mathcal I$ term in Eq.~(\ref{eq:C212aux}).
The third term completes the renormalization of
the $1/L^6$ term in Eq.~(\ref{eq:C212aux}), leaving
a finite residue which we now calculate.

To determine $f'(0)$ we must expand $G$ for small $\vec p$:
\begin{equation}
G(\vec p, \vec q) = f_0 + \vec p\cdot \vec q f_1
+ \vec p^{\,2} f_{2a} + (\vec p\cdot\vec q)^2 f_{2b} + \dots
\,,
\end{equation}
where the $f_j$ are functions of $\vec q^2$.
Performing the angular average over $\vec q$, and picking out
the term quadratic in $\vec p$, we find
\begin{equation}
f'(0) = \int_{\vec q}^\Lambda 
\left( f_{2a}(\vec q^{\,2}) + \frac{\vec q^2}3 f_{2b}(\vec q^{\,2})\right)
\,.
\end{equation}
Carrying out the algebra, and the resulting finite integrals, we find
\begin{equation}
f'(0) = \frac1{m^3} \left( 
\frac1{64\pi}-\frac1{12\pi^2}
- \int_{\vec q}^\Lambda \frac1{4\omega_q^3}\right)
\,.
\end{equation}
The last term is part of the renormalization of the $1/L^6$ term
in Eq.~(\ref{eq:C212aux}), as mentioned above. The first two terms
complete the finite residues at the order we work.

Adding the results from Eqs.~(\ref{eq:C2f}), (\ref{eq:C2fb}),
(\ref{eq:C2i}) and (\ref{eq:C2ib}) we find
\begin{align}
\CT 2 3 &=  \frac{2\mathcal J-\mathcal I^2}{8192 \pi^4 m^4 L^5}
- \frac1{4096 \pi m^5 L^6} 
\nonumber\\
&\quad - \frac1{768 \pi^2 m^5 L^6} + \text{renorm. parts}
\,,
\end{align}
where the ``renorm. parts'' are the above-described contributions
that convert $\lambda_0$ to $\lambda$ in lower-order terms.

With all the results in hand, we can determine $\Delta E_{2,\thr}$
through $\mathcal O(\lambda_0^3/L^6)$. We note that only the first and
third terms in Eq.~(\ref{eq:DeltaE3def}) contribute at this order,
while both terms in Eq.~(\ref{eq:DeltaE2def}) are needed.
Writing the full result in terms of the renormalized coupling
$\lambda$ we obtain
\begin{equation}
\begin{split}
\Delta E_{2,\thr} &= \frac{\lambda}{8m^2L^3} 
- \frac{\lambda^2 \mathcal I}{256 \pi^2 m^3 L^4}
+ \frac{\lambda^3 ({\cal I}^2-{\cal J})}{8192 \pi^4 m^4 L^5} \\
&
\hspace{-20pt} - \frac{3 \lambda^2}{256 m^5 L^6}
+ \frac{\lambda^3}{768 \pi^2 m^5 L^6} + \mathcal O(\lambda^4/L^6,1/L^7)
\,.
\label{eq:DeltaE2thrres}
\end{split}
\end{equation}

\section{Divergence-free three-particle scattering amplitude at threshold}
\label{sec:M3thr}

The threshold energy shift in the three-particle case, $\Delta E_{3,\thr}$, 
depends, at ${\cal O}(1/L^6)$,
on the three-particle scattering amplitude, $\mathcal M_3$.
As is well know, however, $\mathcal M_3$ is singular
for certain choices of external momenta and, in particular, at threshold
(see, e.g., Refs.~\cite{Rubin:1966zz,Brayshaw:1969ab,Taylor:1977A,Taylor:1977B}).
This is a well-understood physical singularity, described, for example
in Ref.~\cite{Hansen:2014eka}. 
It implies that the $1/L^6$ term in $\Delta E_{3,\thr}$ cannot depend on
$\mathcal M_3$ itself, but rather on a subtracted version which is finite at
threshold. In Ref.~\cite{Hansen:2014eka} we provide one possible definition for
a subtracted amplitude, which we call the divergence-free amplitude.
This definition is general, working for particles of any masses and
both at and away from threshold. It is, however, a cumbersome definition
to implement (e.g. involving an infinite number of subtractions, all but
three of which are finite for degenerate particles at threshold). Our analysis of the
threshold expansion of the three-particle quantization condition suggests
instead using a simpler quantity, which we call $\mathcal M_{3,\thr}$,
whose definition involves only the minimum necessary subtractions~\cite{HaSh:inprep}.
This quantity is motivated and discussed at length in Ref.~\cite{HaSh:inprep}. 

The purpose of this section is to calculate $\mathcal M_{3,\thr}$ to
${\cal O}(\lambda^3)$, so that we can express our perturbative result
in terms of this infinite-volume quantity.
Its definition is~\cite{HaSh:inprep}
\begin{align}
\mathcal M_{3,\thr} &\equiv \lim_{\delta\to 0}\Bigg[
{\cal M}_{3,\delta} - I_{0,\delta}
- \int_\delta \frac{d^3 \vec p}{(2\pi)^3} \Xi_1(\vec p)
\nonumber\\
&\qquad\qquad
- \int_\delta \frac{d^3 \vec p_1}{(2\pi)^3}
\int_\delta \frac{d^3 \vec p_2}{(2\pi)^3} \Xi_2(\vec p_1,\vec p_2)\Bigg]
\,.
\label{eq:M3thrdef}
\end{align}
Here $\delta$ is an IR cutoff, whose definition will be explained in the context
of the following calculation. There are three subtraction terms,
involving $I_0$, $\Xi_1$ and $\Xi_2$, respectively. Only the first two terms enter at the order we work, 
since $\Xi_2 = {\cal O}(\lambda^4)$.
We give the definitions of the relevant parts of $I_{0,\delta}$ and $\Xi_1$ below.

\begin{figure}[tbh]
\begin{center}
\includegraphics[scale=0.45]{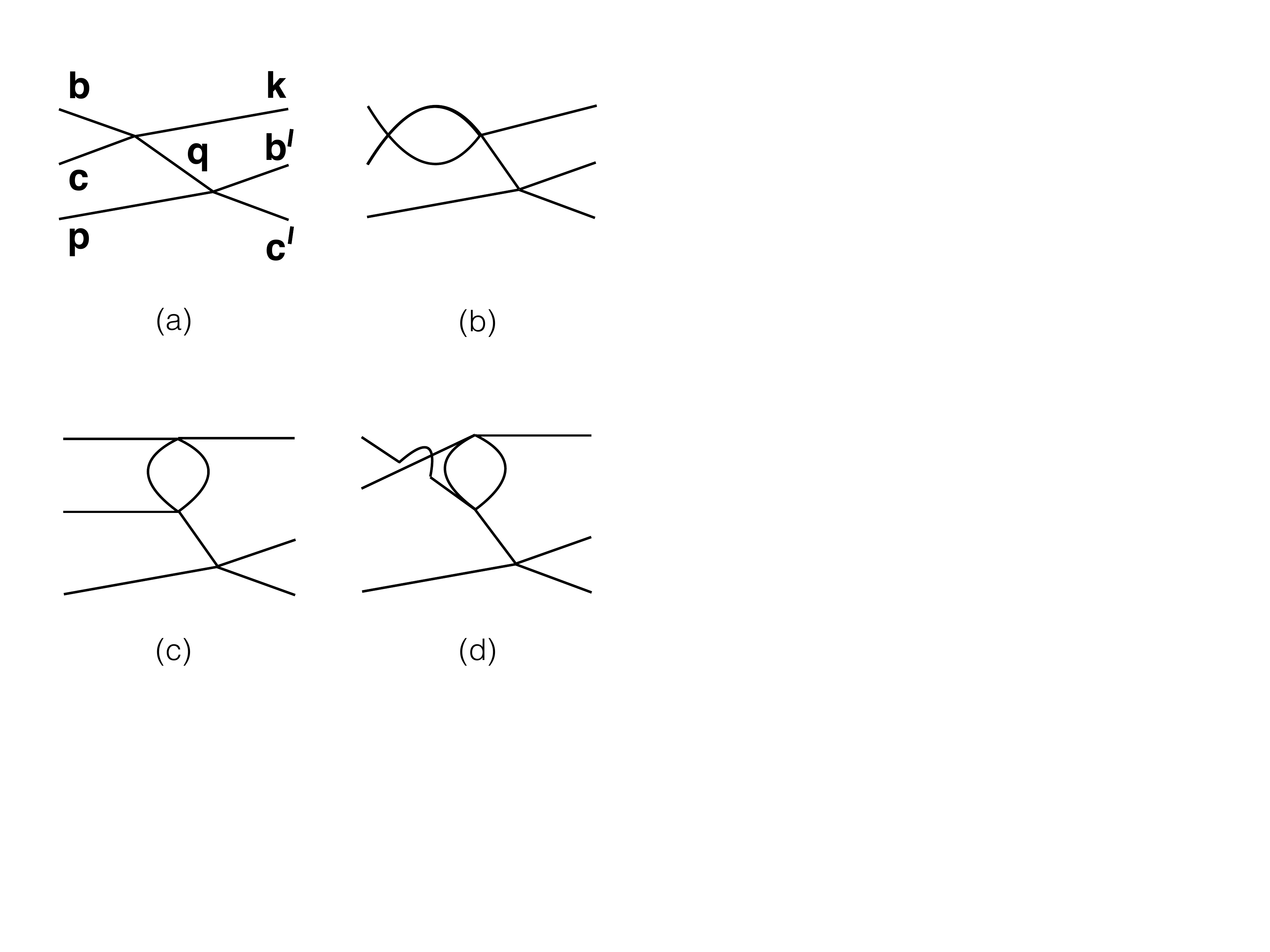}
\vskip -1.3truein
\caption{Feynman diagrams contributing to the three-particle scattering
amplitude, $\mathcal M_3$. The momentum labels shown in (a) apply
to all four diagrams. }
\label{fig:treeloop}
\end{center}
\end{figure}

The diagrams that contribute to $\mathcal M_3$ at the order we work are
those of Figs.~\ref{fig:C3}(i)-(o), now interpreted as Feynman diagrams 
in infinite volume. We first consider the ${\cal O}(\lambda_0^2)$ diagram,
Fig.~\ref{fig:C3}(i), which is also reproduced in
Fig.~\ref{fig:treeloop}(a) along with the momentum labels we use.
This gives a contribution that
diverges at threshold, since the intermediate propagator goes on shell.
The $\delta$ prescription of Eq.~(\ref{eq:M3thrdef}) corresponds here simply
to working away from threshold with general momenta,
making the subtraction (here of $I_0$),
and then sending all external momenta to zero~\cite{HaSh:inprep}.
The contribution to the scattering amplitude is 
\begin{align}
\mathcal M_3^{(i)} &= -\frac{\lambda_0^2}{q^2-m^2 +i\epsilon}
\,,\label{eq:M3i}
\\
q & = (E - \omega_p-\omega_k, -\vec p -\vec k)
\end{align}
where $ E=\omega_p+\omega_a+\omega_b$, and we have
used the vanishing of the total spatial momentum.
%At this order there is no dependence on the directions of $\vec a$ and $\vec b$,
%or their final-state counterparts.
The denominator of the propagator is
\begin{equation}
q^2-m^2 = (E- W)(E-W+2\omega_{kp})
\,,
\end{equation}
where $\omega_{kp}=\sqrt{(\vec p\!+\!\vec k)^2+m^2}$ and,
here, $W=\omega_p\!+\!\omega_k\!+\!\omega_{kp}$.
This denominator vanishes when all the spatial momenta tend to zero, for then
$E$ and $W$ both tend to $3m$. 
It also vanishes for non-zero momenta
if $E=W$ (i.e. if $\omega_a+\omega_b=\omega_k+\omega_{kp}$) and we must avoid
such above-threshold divergent momentum configurations.
The subtraction term corresponding to this diagram is the leading order
part of $I_0$~\cite{HaSh:inprep}
\begin{equation}
I_0^{(i)} = - \frac{\lambda_0^2}{2 \omega_{kp} (E-W +i \epsilon)}
\,.
\label{eq:I0i}
\end{equation}
By construction this has the same pole at $E=W$ as $\mathcal M_3^{(i)}$, 
so that the difference is finite (both at threshold and for above-threshold divergent momenta):
\begin{equation}
\mathcal M_3^{(i)}-I_0^{(i)} = \frac{\lambda_0^2}{2\omega_{kp}(E-W+2\omega_{kp})}
\,.
\end{equation}
The final steps needed to obtain $\mathcal M_{3,\thr}$ are to symmetrize over the
external momentum assignments,\footnote{%
Since the particles are identical, interchanging $\vec a$ and $\vec b$ does not
lead to a different assignment, so there are only three choices for each of the
initial and final states}
and to take the threshold limit.
Since the limiting value is independent of the choice of external momenta, symmetrization
gives a factor or $3\times 3$, and the total contribution from this diagram is
\begin{equation}
\mathcal M_{3,\thr}^{(i)} = \frac{9\lambda_0^2}{4m^2}
\,.
\label{eq:M3thri}
\end{equation}

We next consider the diagrams in which the vertices in the tree diagram
receive one-loop corrections. 
The full set of of these are those involving the s-, t- and u-channel loops
shown in Figs.~\ref{fig:treeloop}(b), (c) and (d), respectively,
as well as the corresponding corrections to the right-hand vertex.
The subtraction in this case is somewhat subtle so we provide 
a more detailed explanation.

A key point in the following is that the intermediate propagator
(with momentum $q$) is off shell, and so the loop corrections
to the vertices at either end of this propagator differ in general from those
which enter the on-shell scattering amplitude. This is true despite
the fact that three of the four legs are on shell (e.g. those with
momenta $\vec b$, $\vec c$ and $\vec k$ for the left-hand vertex).
In fact, at one-loop order, the s-channel loops [e.g. Fig.~\ref{fig:treeloop}(b)] 
do not depend on the off-shellness of $q$, while the t and u-channel loops 
[e.g. Figs.~\ref{fig:treeloop}(c) and (d)] do.

This point is important because the general form of the subtraction term,
$I_0$, replaces the loops at either end of the central propagator with on-shell
scattering amplitudes, as well as changing the form of the propagator (as described
above for the tree-level diagram). This feature of the subtraction is crucial, since it
means that it is given in terms of physical quantities.
The general form is quite complicated because it involves scattering amplitudes
in all partial waves~\cite{HaSh:inprep}. However, since we are interested in the
subtracted amplitude at threshold, we need only the part of $I_0$ that involves s-wave
scattering, and we need this only close to threshold. 
Specifically, we have
\begin{align}
 I_0^{(ijk)} &= - 
 \frac{{\cal M}_2(b^*) {\cal M}_2(b'^*)}{2\omega_{kp}(E-W+i\epsilon)}
 \,,
 \label{eq:I0ijk}
 \\ \begin{split}
 {\cal M}_2(b) &=
- \lambda  - \frac{\lambda^2}{4 \pi}\left( -\frac{i b}{8 m}
 + \frac{b^{2}}{3 m^2} \right) %\\[5pt] & \hspace{130pt} 
 + \mathcal O(b^3, \lambda^3)
% {\cal M}_2(a) &=
%  \left\{\lambda + {\lambda^2}\left( -\frac{i a}{32\pi \omega_{a}}
% + \frac{a^{2}}{12\pi m^2} \right) \right\}
 \,, \label{eq:M2expand} \end{split}
 \end{align}
where the superscript ``$(ijk)$'' refers to the subfigures within Fig.~\ref{fig:C3}. 
Here $b^*$ is the magnitude of the spatial momentum of $b$ evaluated
 in the $b+c$ CM frame, and $b'^*$ is the same quantity for $b'$. 
 The result (\ref{eq:M2expand}) is simply the threshold expansion of
 the two-particle scattering amplitude, obtained using Eqs.~(\ref{eq:B3}),
 (\ref{eq:lambdadef}) and (\ref{eq:B13}) in Appendix~\ref{app:effectiverange}.
 Note that this is expressed in terms of $\lambda=\lambda_0(1 - A_2 \lambda_0 + \dots)$.
 Note also that the leading order part of Eq.~(\ref{eq:I0ijk}) is the subtraction
 (\ref{eq:I0i}) that we used for the tree-level diagram.
 
 Evaluating Figs.~\ref{fig:treeloop}(b), (c) and (d), together with the diagrams where the
 other vertex is corrected, and adding the result (\ref{eq:M3i}) from 
 Fig.~\ref{fig:treeloop}(a), we obtain
 \begin{multline}
 {\cal M}_3^{(ijk)} = -
 \left(\mathcal M_2(b^*)+\frac{\lambda_0^2(q^2-m^2)}{192 \pi^2 m^2}\right)
 \\
\times\frac1{q^2-m^2+i\epsilon}
 \left(\mathcal M_2(b'^*) + \frac{\lambda_0^2(q^2-m^2)}{192 \pi^2 m^2} \right)
 \,.
 \end{multline}
As noted above, only the t- and u-channel loops differ from the on-shell scattering amplitude.
The part which differs is, close to threshold, proportional to $t+u$, 
where, for example, $t=(b-k)^2$ and $u=(c-k)^2$ when
the left-hand vertex is being corrected. Now $t+u = q^2+3 m^2 -s$, and
$s=(b+c)^2$ is the same irrespective of whether $q$ is on or off shell.
Thus the difference between $t+u$ on- and off-shell is equal to the difference 
between $q^2$ on and off-shell, which is just $q^2 - m^2$. 
The key point is that this cancels the denominator of the
propagator, leaving a finite residue in the threshold limit.

We can now perform the subtraction, take the threshold limit, and multiply by
the momentum permutation factor of 9, to obtain the final result from
the diagrams of Fig.~\ref{fig:treeloop}:
\begin{equation}
\mathcal M_{3,\thr}^{(ijk)} = \frac{9 \lambda^2}{4m^2} + \frac{9 \lambda^3}{96 \pi^2 m^2}
+{\cal O}(\lambda^4)
\,.
\end{equation}
Note that we have written this result in terms of the renormalized coupling.

The ``bull's head'' diagram of Fig.~\ref{fig:C3}(l) also requires subtraction,
in this case from the $\Xi_1$ term in Eq.~(\ref{eq:M3thrdef}).
At ${\cal O}(\lambda_0^3)$ the subtraction function is
\begin{equation}
\Xi_1(\vec p) = - \frac{9 \lambda_0^3}{8m} \frac{H(\vec p)^2}{(\vec p^2)^2}
\,,
\label{eq:Xi13}
\end{equation}
where $H$ is the UV cutoff function discussed in Appendix~\ref{app:sums}.
$\Xi_1$ is to be integrated over $\vec p$ with an IR cutoff $|\vec p \,| \ge \delta$,
so that the integral is finite.
The definition of the IR-regulated three-particle scattering amplitude,
$\mathcal M_{3,\delta}$ is, in general, quite involved, as it requires
applying a cut-off both on loop momenta and also on the external momenta.
However, the external momenta scale as $\delta^{3/2}$, which implies
that the IR cutoff they induce in the loop integral is weaker than 
that from the direct cutoff at $|\vec p \,|=\delta$. This means that we can
set the external momenta to zero. Doing so, we find that
\begin{equation}
\mathcal M_{3,\delta}^{(l)} = -\frac{9\lambda_0^3}{16}
\int_{\vec p,\delta} \frac{3\omega_p^2-m^2}{\omega_p^3 (\vec p^2)^2}
\,,
\label{eq:M3deltal}
\end{equation}
where the subscript $\delta$ on the integral indicates the IR cutoff.
Combining $\mathcal M_{3,\delta}^{(l)}$ with the integral of $\Xi_1$
leads to an IR convergent integral in which one can set $\delta=0$:
\begin{equation}
\mathcal M_{3,\thr}^{(l)} = - \frac{9\lambda_0^3}{16 m^2} \int_{\vec p} 
\left[\frac{3\omega_p^2-m^2}{\omega_p^3}-\frac{2H(\vec p)^2}{m}\right]
\frac{m^2}{(\vec p^2)^2}
\,.
\label{eq:M3thrl}
\end{equation}
We stress that this is a UV convergent integral that has a finite value,
but that this value depends on the choice of cutoff function $H$.

Next we consider the ``twisted bull's head" diagram of Fig.~\ref{fig:C3}(m). This diagram is
convergent in the IR, so no subtraction is needed, and it can be evaluated directly
at threshold. We find 
\begin{align}
{\cal M}_{3,\thr}^{(m)} &= -i 6 \lambda_0^3 \int \frac{d^4p}{(2\pi)^4} 
\left(\frac1{p^2-m^2+i\epsilon}\right)^3 \,,
\\
& = - \frac{3\lambda_0^3}{16 \pi^2 m^2}
\,.
\label{eq:M3thrm}
\end{align}

The final diagrams involve intermediate single-particle states. These also have no IR
divergences, require no subtractions, and can be evaluated directly at threshold.
Since Fig.~\ref{fig:C3}(n) is a tree diagram, it is simple to evaluate, and yields
\begin{equation}
{\cal M}_{3,\thr}^{(n)} = - \frac{\lambda_0^2}{8 m^2}
\,.
\label{eq:M3thrn}
\end{equation}
The one-loop corrections to this diagram are given by
Fig.~\ref{fig:C3}(o) and the similar diagram where the right-hand vertex is corrected.
Note that although this looks like an s-channel loop and might be expected to have
t- and u-channel partners, in fact, no other diagrams exist. 
Evaluating the loop we find that the sum of the two diagrams gives
\begin{equation}
{\cal M}_{3,\thr}^{(o)} = \frac{3\lambda_0^3}{32 m^2} \int_{\vec p}^\Lambda \frac1{\omega_p \vec p^2}
\,.
\label{eq:M3thro}
\end{equation}
The UV divergence here is exactly that needed to convert $\lambda_0$ in (\ref{eq:M3thrn})
to $\lambda$, up to a finite residue:
\begin{equation}
{\cal M}_{3,\thr}^{(no)} = - \frac{\lambda^2}{8m^2} + \frac{\lambda^3}{32\pi^2m^2} 
+{\cal O}(\lambda^4)
\,.\label{eq:M3thrn+o}
\end{equation}

% SS:new material
\bigskip
Combining results from all diagrams, we obtain
\begin{equation}
\mathcal M_{3,\thr} = \frac{9\lambda^2}{4m^2} + \frac{3\lambda^3}{32\pi^2 m^2}
+ \mathcal M_{3,\thr}^{(lmno)} + {\cal O}(\lambda^4)
\,,
\label{eq:Mthrtot}
\end{equation}
where the last quantity is the contribution from Figs.~\ref{fig:C3}(l)-(o),
given by
\begin{align}
\mathcal M_{3,\thr}^{(lmno)} &=
- \frac{\lambda^2}{8m^2}+ \frac{\lambda^3}{8 \pi^2 m^2}
+ \frac{9 \lambda^3 {\mathcal C}_3}{8 m^2} \,,
\label{eq:M3thrlmno}
\end{align}
with
\begin{equation}
\mathcal C_3 = \int_{\vec p} \frac{m [H(\vec p)^2-1]}{(\vec p^2)^2}
\,.
\end{equation}
We observe that, at the order we work, one could dispense with
the cutoff function $H$, i.e. set $H=1$ so that $\mathcal C_3=0$.
This is appealing because it would remove the cutoff dependence
from $\mathcal M_{3,\thr}$ at this order.
However, this removal does not extend to higher orders in $\lambda$. 

\section{Three particle energy shift}
\label{sec:C3}

In this section we determine the three-particle threshold energy shift,
which (aside from renormalization) arises from the 
diagrams shown in Fig.~\ref{fig:C3}. The fully disconnected diagram,
Fig.~\ref{fig:C3}(a), was discussed in Sec.~\ref{sec:methods},
and leads to the leading ``$1$'' in $C_{3,\thr}^{(0)}$,
Eq.~(\ref{eq:Cjthr0exp}). The next seven diagrams,
Figs.~\ref{fig:C3}(b)-(h), involve the interaction of only two particles,
with the third being a spectator. Thus the results are the same as
those for the corresponding two-particle diagrams,
discussed in Sec.~\ref{sec:C2}, except multiplied by a factor of three
for the number of pairs. Calling these contributions ``disconnected'',
we thus have, for the part linear in $\tau$,
\begin{equation}
\partial_\tau C_{3,\thr,\disc}^{(n)}(0) = 3\; \partial_\tau C_{2,\thr}^{(n)}(0) \,, \ \ \  n=1,2,3,\dots
%C_{3,\thr,\disc}^{(1,n)} = 3\; C_{2,\thr}^{(1,n)}\,,\qquad n=1,2,3,\dots
\label{eq:C1ndisc}
\end{equation}
For the constant term, where we need only the terms linear and quadratic in 
$\lambda_0$, and only out to $1/L^3$ in the volume expansion,
it turns out that ``connected'' diagrams do not contribute.
Thus the rescaled two-particle results are all that we need:
\begin{equation}
\CZ 3 n = 3\; \CZ 2 n \,,\qquad n=1 \ {\rm and}\ 2 \,.
\label{eq:C0ndisc}
\end{equation}
Our remaining task is therefore to calculate the connected contributions to 
$\CTT 3$, which arise from Figs.~\ref{fig:C3}(i)-(o).

We begin with the tree diagram Fig.~\ref{fig:C3}(i). A simple calculation
results in\footnote{%
There is also contribution to $\CZ 3 2$ proportional to
$1/L^6$, but this is beyond the order we need, as noted above.}
\begin{equation}
\CT 3 2 \supset \frac{3}{32 m^5 L^6}
\,.
\label{eq:C3thr12i}
\end{equation}
One-loop corrections to this result are given by
Figs.~\ref{fig:C3}(j) and (k) and their reflections.
The contributions from vanishing loop momenta scale as $1/L^9$ and
can be dropped. Those from non-vanishing loop momenta give, 
after combining all four diagrams,
\begin{align}
\CT 3 3 &\supset - \frac3{256 m^5 L^6} S^{(jk)}
\,,
\\
S^{(jk)} = \frac1{L^3} & \!\sum_{\vec p}
\left[ \frac2{\omega_p p^2}\!+\! \frac4{\omega_p^3}
 \!-\! \frac{m^2}{\omega_p^4(\omega_p+m)}\!-\! \frac{2 m^2}{\omega_p (\vec p^2)^2}
\right]
\,.
\end{align}
For the first three summands we can replace the sum with an integral
at the order we work, while for the last we must use Eq.~(\ref{eq:SJres}).
We also note that the first two summands are proportional to the integrand
of $A_2$ [see Eq.~(\ref{eq:A2def})], and indeed have the correct
normalization to convert $\lambda_0^2$ in 
Eq.~(\ref{eq:C3thr12i}) to $\lambda^2$.
Collecting all contributions and carrying out the resulting finite integrals,
we find
\begin{equation}
S^{(jk)} \supset  - \frac{mL {\cal J}}{8\pi^4} + 16 A_2 + \frac1{2\pi^2}
+\frac1{8\pi}
\,.
\end{equation}

Next we consider the bull's head diagram, Fig.~\ref{fig:C3}(l).
Again the $\vec p=0$ contribution  can be dropped.
For non-zero loop momentum we find
\begin{align}
\CT 3 3 &\supset - \frac{3}{256 m^5 L^6} S^{(l)}
\,,\\
S^{(l)} 
&= \frac1{L^3} \sum_{\vec p\ne 0} 
\frac{m^2(3\omega_p^2-m^2)}{\omega_p^3 (\vec p^2)^2}
\,.
\end{align}
Using Eq.~(\ref{eq:SJres}) we can write the sum as
\begin{equation}
S^{(l)} 
= \frac{mL}{8\pi^4} {\cal J}
+ \int_{\vec p}
\left\{\frac{(3\omega_p^2-m^2)}{\omega_p^3}-\frac2{m}\right\} 
\frac{m^2}{(\vec p^2)^2} +{\cal O}(L^{-1})
\,.
\end{equation}
The integral appearing here is essentially the same as that in the
subtracted bull's head contribution to the scattering amplitude,
Eq.~(\ref{eq:M3thrl}), except that the cutoff function $H$ is absent.
Combining results we find the total bull's head contribution to be
\begin{align}
\CT 3 3 &\supset - \frac{3 {\cal J}}{2048 \pi^4 m^4 L^5}
+ \frac{\mathcal M_{3,\thr}^{(l)}}{48 \lambda_0^3 m^3  L^6}
%\nonumber\\
%&\qquad 
- \frac{3 {\cal C}_3}{128 m^5 L^6} 
%\int_{\vec p} \frac{m[1-H(\vec p)^2]}{(\vec p^2)^2}
\,.
\end{align}

The twisted bull's head diagram of Fig.~\ref{fig:C3}(m) contributes
\begin{align}
\CT 3 3 &\supset -\frac{3}{128 m^5 L^6} 
\frac1{L^3}\! \sum_{\vec p} \frac{m^2}{\omega_p^5} \,,
\\
&= - \frac{1}{256 \pi^2 m^5 L^6}
\,,
\\
&= \frac{\mathcal M_{3,\thr}^{(m)}}{48 \lambda_0^3 m^3 L^6}
\,,
\end{align}
where in the second line we have converted the IR and UV finite 
sum to an integral and evaluated the integral,
while in the third line we have used the result for the corresponding contribution
to the scattering amplitude, Eq.~(\ref{eq:M3thrm}).

Finally we come to the diagrams involving single-particle intermediate states,
Figs.~\ref{fig:C3}(n) and (o), together with the reflection of the latter.
Since the presence of growing exponentials is a new feature of these diagrams,
we give a little more detail here. The tree diagram of Fig.~\ref{fig:C3}(n) leads to
\begin{equation}
C_{3}(\tau) \supset \frac{\lambda_0^2}{96 m^6 L^6}
\left(\frac9{16} e^{2m\tau} - \frac{13}{48} - \frac{m \tau}{4}\right)
\,.
\end{equation}
Dropping the enhanced exponential, the $\tau$ term gives the contribution
\begin{align}
\CT 3 2 &\supset - \frac1{384m^5 L^6} 
= \frac{\mathcal M_{3,\thr}^{(n)}}{48  \lambda_0^2 m^3L^6} 
\,,
\end{align}
where we have used the result Eq.~(\ref{eq:M3thrn}) for the contribution to
the threshold amplitude from Fig.~\ref{fig:C3}(n).

This by-now standard relation between the threshold amplitude and the $1/L^6$ energy
shift holds also for the one-loop diagram Fig.~\ref{fig:C3}(o). 
Including its reflection we find the contribution
\begin{equation}
\CT 3 3\supset \frac{1}{512 m^5 L^6} 
\frac1{L^3}\!\sum_{\vec p\ne 0} \frac1{\omega_p \vec p^2}
\,.
\end{equation}
Since the sum can be replaced by an integral up to ${\cal O}(1/L^7)$ corrections,
we see from Eq.~(\ref{eq:M3thro}) that this contribution 
indeed equals $\mathcal M_{3,\thr}^{(o)}/(48 \lambda_0^3 m^3 L^6)$.

% SS: new material
\bigskip
The total result from all connected diagrams is
\begin{multline}
\lambda_0^2 \big [ \partial_\tau C_{3,\thr,\conn}^{(2)}(0) \big ] + \lambda_0^3 \big [ \partial_\tau C_{3,\thr,\conn}^{(3)}(0) \big ] 
= 
\\
\frac{3 \lambda^2}{32 m^5 L^6}
- \frac{3 \lambda^3}{512\pi^2 m^5 L^6}
- \frac{3 \lambda^3}{2048 \pi m^5 L^6}
\\
- \frac{3 \lambda^3 {\mathcal C_3}}{128 m^5 L^6}
+ \frac{{\cal M}_{3,\thr}^{(lmno)}}{48 m^3 L^6}
+ {\cal O}(\lambda^4) \,,
\end{multline}
where 
$\mathcal M_{3,\thr}^{(lmno)}$ is given in Eq.~(\ref{eq:M3thrlmno}).
We observe that the $\lambda_0^3 {\cal J}/L^5$ terms cancel between
Figs.~\ref{fig:C3}(j), (k) and (l).
Combining this with the results of Eqs.~(\ref{eq:C1ndisc})
and (\ref{eq:C0ndisc}) for the disconnected diagrams, and using the relation
(\ref{eq:Mthrtot}) between $\mathcal M_{3,\thr}$ and
$\mathcal M_{3,\thr}^{(lmno)}$, 
together with expressions (\ref{eq:DeltaE1def})-(\ref{eq:DeltaE3def})
for the energy shift, we obtain
\begin{multline}
\Delta E_{3,\thr} =
\frac{3\lambda}{8m^2 L^3}
- \frac{3\lambda^2\mathcal I}{256\pi^2 m^3 L^4}
\\
+ \frac{3 \lambda^3(\mathcal I^2+\mathcal J)}{8192 \pi^4 m^4 L^5}
- \frac{9\lambda^2}{256 m^5 L^6}
+ \frac{3\lambda^3}{256\pi^2 m^5 L^6}
\\
+ \frac{3 \lambda^3 \mathcal C_3}{128 m^5 L^6}
- \frac{\mathcal M_{3,\thr}}{48 m^3 L^6}
+ \mathcal O(\lambda^4/L^6, 1/L^7) \,.
\end{multline}
This is the main result of this article. 
We recall that $-\lambda$ is defined to equal the threshold two-to-two scattering amplitude,
and is related to the scattering length via $\lambda = 32 m \pi a$.
Finally we observe that the dependence on the cutoff function
$H$ cancels between the last two terms, as must be the case
since the finite-volume energy shift is a physical quantity.

\section{Comparisons and conclusions}
\label{sec:conc}

To compare our results for the threshold energy shifts to those in the
literature, it is convenient to re-express them in terms of the
scattering length.
For the two-particle case we obtain
\begin{align}
\Delta E_{2,\thr}\! &=\! \frac{4\pi a}{m L^3} \left( 1
\!-\! \frac{a \mathcal I}{\pi L}
\!+\! \frac{a^2 (\mathcal I^2\!-\!\mathcal J)}{\pi^2 L^2}\right)
\!+\! \frac{a_2^{(6)}}{L^6} \!+\! \mathcal O(L^{-7})
\,,\\
a_2^{(6)} &= \frac{4\pi a}{m} \left(\!-\frac{3\pi a}{m^2}
\!+\! \frac{32 a^2}{3 m}\right) + \mathcal O(a^4)
\,.
\label{eq:a6intermsofa}
\end{align}
The $1/L^3$, $1/L^4$ and $1/L^5$ contributions agree with those obtained
in Refs.~\cite{Luscher:1986n2,Beane2007} (Reference~\cite{Tan2007} did not
consider this quantity.)
To aid comparison between the $1/L^6$ terms, we rewrite $a_2^{(6)}$
in terms of the 
effective range $r$ given in Eq.~(\ref{eq:effectiverange}):
\begin{align}
a_2^{(6)} &= 
%-\frac{12 \pi^2 a^2}{m^3} + \frac{128 \pi a^3}{3 m^2} + \mathcal O(a^4)
%\\ &= 
-\frac{4 \pi^2 a^2}{m^3} + \frac{8 \pi^2 a^3 r}{ m}+ \mathcal O(a^4)
%\frac{4\pi a}{m} \left(
% -\frac{\pi a}{m^2} + {2 \pi a^2 r}\right) + \mathcal O(a^4)
\,.
\label{eq:a6intermsofr}
\end{align}
The latter form agrees with that obtained in Appendix~\ref{app:Luscher}
from expanding L\"uscher's quantization condition~\cite{Luscher:1991n1} 
to ${\cal O}(1/L^6)$. 
It is, however, in disagreement with 
\begin{equation}
a_2^{(6)}(\text{Ref.~\cite{Beane2007}})=
\frac{8 \pi^2 a^3 r}{m} + {\cal O}(a^4)\,.
\end{equation}
The disagreement, which is proportional to $\lambda^2/L^6$ in $\Delta E_{2,\thr}$,
appears to arise from the use of a non-relativistic dispersion relation
at the last stage of the calculation, as discussed in Appendix~\ref{app:Luscher}. We stress that $ a^3 r$ starts at $\mathcal O(\lambda^2)$ in PT,
so that the result from Ref.~\cite{Beane2007}  does contain a term scaling as $\lambda^2/L^6$, 
but with a different coefficient from that which we find here.

We note that L\"uscher's quantization condition holds for general scalar field theories, including effective
field theories with arbitrary higher-order couplings. Thus the form of $a_2^{(6)}$ that holds in all
such theories is that given in Eq.~(\ref{eq:a6intermsofr}), i.e. in terms of $a$ and $r$.
By contrast, the form given in Eq.~(\ref{eq:a6intermsofa}) holds only for $\lambda \phi^4$ theory.

Our result for the three-particle energy shift is
\begin{align}
\Delta E_{3,\thr}\! &=\! \frac{12\pi a}{m L^3} \left( 1
\!-\! \frac{a \mathcal I}{\pi L}
\!+\! \frac{a^2 (\mathcal I^2\!+\!\mathcal J)}{\pi^2 L^2}\right)
\!+\! \frac{a_3^{(6)}}{L^6} \!+\! \mathcal O(L^{-7})
\,,\label{eq:final1}
\end{align}
where
%\begin{comment}
\begin{align}
\hspace{-10pt} a_3^{(6)} &= 
\frac{12\pi a}{m} \left(\!-\frac{3\pi a}{m^2}
\!+\! \frac{32 a^2}{m}\right)
\!+\! \frac{768 a^3 \pi^3 \mathcal C_3}{m^2}
\!-\! \frac{{\cal M}_{3,\thr}}{48 m^3} 
\label{eq:final2} \,,  \\
&=
\frac{12\pi a}{m} \left(\frac{3\pi a}{m^2} + 6\pi a^2 r \right)
\!+\! \frac{768 a^3 \pi^3 \mathcal C_3}{m^2}
\!-\! \frac{{\cal M}_{3,\thr}}{48 m^3} 
\,.
\label{eq:final3}
\end{align}
%\end{comment}
%\begin{multline}
%a_3^{(6)} = \frac{36\pi^2 a^2}{m^3} + \frac{72\pi^2 a^3 r }{m}
%\\\!+\! \frac{768 a^3 \pi^3 \mathcal C_3}{m^2}
%\!-\! \frac{{\cal M}_{3,\thr}}{48 m^3} + \mathcal O(a^4)
%\label{eq:final3}\,,
%\end{multline}
As for the two particle case, 
the second expression holds for general interactions whereas that in terms of $a$ alone 
is special to $\lambda \phi^4$ theory.
The $1/L^3$, $1/L^4$ and $1/L^5$ contributions agree with those obtained
in Refs.~\cite{Beane2007,Tan2007}.
As for $a_3^{(6)}$, Ref.~\cite{Beane2007} finds
\begin{equation}
a_3^{(6)}(\text{Ref.~\cite{Beane2007}})=
\frac{24 \pi^2 a^3 r}{m} + \eta_3(\mu) + {\cal O}(a^4)\,,
\label{eq:Beane3}
\end{equation}
while Ref.~\cite{Tan2007} quotes
\begin{equation}
a_3^{(6)}(\text{Ref.~\cite{Tan2007}})=
\frac{36 \pi^2 a^3 r}{m} + D + {\cal O}(a^4)\,.
\label{eq:Tan3}
\end{equation}
Here $\eta_3(\mu)$ is a three-particle contact interaction
while $D$ is the ``three-body scattering hypervolume''.
Both characterize a local three-particle interaction within
the non-relativistic context of their respective calculations.
The scale $\mu$ in $\eta_3$ is a renormalization scale, and any dependence
on this scale must cancel out in the energy shift.
The results of Ref.~\cite{Beane2007}
show, however, that this dependence enters only at ${\cal O}(a^4)$.

While $\eta_3$ and $D$ are non-relativistic analogs of our threshold
amplitude $\mathcal M_{3,\thr}$, there could, in general, be finite differences
between these quantities. Indeed equating the three results in
Eqs.~(\ref{eq:final3}), (\ref{eq:Beane3}) and (\ref{eq:Tan3}) gives
relations between these quantities.
However, given that our relativistic calculation gives
a different result for $a_2^{(6)}$ from that obtained
using non-relativistic methods in Ref.~\cite{Beane2007}, it is not
clear whether the non-relativistic three particle results apply at
${\cal O}(1/L^6)$ in a relativistic theory.

Irrespective of these considerations, our result allows us to check the
threshold expansion that we obtain in Ref.~\cite{HaSh:inprep}
from our relativistic three-particle
quantization condition~\cite{Hansen:2014eka,Hansen:2015zga}.
We find complete agreement, giving us further confidence in the
quantization condition.
Furthermore, the perturbative calculation carried out here
has provided a first explicit verification of the details of
the subtractions needed to define a finite three-particle scattering
amplitude at threshold.

It would be interesting to push this calculation to one higher order in
$\lambda$, so as to allow a check of the $\lambda^4/L^6$ terms that
arise in the threshold expansion of Ref.~\cite{HaSh:inprep}.

\section*{Acknowledgments}
The work of SS was supported in part by the United States Department of Energy 
grant DE-SC0011637. We thank Silas Beane and Martin Savage for discussions and comments.

%\newpage

\appendix

\section{Identities for finite-volume sums}
\label{app:sums}

We collect here various results needed in the main text to convert sums into
integrals plus a finite-volume residue. Throughout we drop terms that are suppressed
exponentially, i.e.~as $\exp(-mL)/L^n$ for $n\ge 0$.
We make use of a cutoff function $H(\vec p\,)$ that was introduced in
Ref.~\cite{Hansen:2014eka}.\footnote{%
$H(\vec p\,)$ is actually a function of $\vec p^{\,2}$ but we use  $\vec p$ as the argument for brevity.
We note that the derivation can be made without introducing $H$, but doing so allows us to
borrow a result from Ref.~\cite{Hansen:2014eka}.}
The relevant properties of $H$ are that it equals unity for $\vec p^{\,2}=0$,
vanishes for $\vec p^{\,2} \ge 16 m^2/9$, and interpolates smoothly in between.
In addition, all derivatives of $H$ at $\vec p^{\,2}=0$ and at $\vec p^{\,2}=16 m^2/9$ vanish. 

We first consider the generic ``$1/\vec p^{\,2}$" sum 
\begin{align}
S_I &= \frac1{L^3} \sum_{\vec p\ne 0}^\Lambda \frac{f(\vec p^2)}{\vec p^2}\,.
\end{align}
As always in this appendix, the sum is over finite-volume momenta,
in this case excluding $\vec p=0$ where the summand diverges.
The function $f(\vec p^{\,2})$ is assumed regular at $\vec p=0$, so that the
corresponding integral is IR convergent in three dimensions.
In general, however, the sum is UV divergent, and must be regularized in some way,
as indicated by the superscript $\Lambda$ on the sum.
All we need to know about this regularization is that it involves a cutoff
scale $\Lambda \gg 4 m/3$, so that $H=0$ at the cutoff scale.

\begin{comment}

{\mh To proceed, we rewrite the sum as
\begin{align}
S_I &= \int_{\vec p}^\Lambda \frac{f(\vec p^2)}{\vec p^2} + \bigg [ \frac1{L^3} \sum_{\vec p\ne 0}^\Lambda -  \int_{\vec p}^{\Lambda} \bigg ] \frac{f(\vec p^2)}{\vec p^2} \,.
\end{align}
In the second term we then substitute $f(\vec p^{\,2}) = f(0) + f'(0) \vec p^{\,2} + \mathcal O(\vec p^{\,4})$ to reach
\begin{multline}
S_I = \int_{\vec p}^\Lambda \frac{f(\vec p^2)}{\vec p^2} + f(0)  \bigg [ \frac1{L^3} \sum_{\vec p\ne 0}^\Lambda -  \int_{\vec p}^{\Lambda} \bigg ] \frac{1}{\vec p^2} \\
 - \frac{f'(0)}{L^3}  + \bigg [ \frac1{L^3} \sum_{\vec p}^\Lambda -  \int_{\vec p}^{\Lambda} \bigg ] \frac{f(\vec p^2)-f(0)}{\vec p^2}  \,.
\end{multline}
In the second line we have subtracted and added the $\vec p=0$ term. This is convenient because the last term, with $\vec p=0$ included, is exponentially suppressed in $m L$. Substituting an identity for the factor multiplying $f(0)$ on the first line, we conclude
\begin{equation}
S_I = \int_{\vec p}^\Lambda \frac{f(\vec p^2)}{\vec p^2} + \frac{\mathcal I f(0)}{4 \pi^2 L} - \frac{f'(0)}{L^3} + \mathcal O(e^{- m L}) \,,
\end{equation}
where ${\cal I}$ is a geometrical factor, equal to $Z_{00}(1,0)$
of Ref.~\cite{Luscher:1986n2}.
}

\end{comment}

To proceed, we rewrite the sum as\footnote{%
The appearance of $H^2$ in the extra terms
is simply to match the results that appear in the threshold expansion of
the three-particle quantization condition~\cite{HaSh:inprep}.
For the derivation we could equally well use $H$ alone.}
\begin{align}
S_I &= \frac1{L^3} \sum_{\vec p\ne 0}^\Lambda 
\left( \frac{f(\vec p^2)}{\vec p^2} - \frac{f(0) H(\vec p)^2}{\vec p^2}\right)
\nonumber\\
&\quad+ \bigg[ \frac{1}{L^3}\sum_{\vec p\ne 0}^\Lambda - \int_{\vec p}^\Lambda \bigg ] 
\frac{f(0) H(\vec p)^2}{\vec p^2}
\nonumber\\
&\quad + \int_{\vec p}^\Lambda \frac{f(0) H(\vec p)^2}{\vec p^2}\,.
\label{eq:SIv2}
\end{align}
%where $\int_{\vec p} \equiv \int d^3p/(2\pi)^3$. 
Given the properties of $H$, the combined summand on the first line is
regular at $\vec p=0$. Thus, if one adds the $\vec p=0$ term to the sum,
it can be replaced by an integral up to exponentially suppressed corrections.
Thus the first line becomes
\begin{equation}
\int_{\vec p}^\Lambda \left( \frac{f(\vec p^2)}{\vec p^2} - \frac{f(0) H(\vec p)^2}{\vec p^2}\right)
- \frac{f'(0)}{L^3} + \mathcal O(e^{-mL})\,.
\end{equation}
In the sum-integral difference on the second line of Eq.~(\ref{eq:SIv2}) the cut-off
$\Lambda$ can be dropped since the summand-integrand is regulated 
in the UV by $H^2$. This difference is then nothing other than a regulated
$1/\vec p^{\,2}$ sum. Using a result derived in Ref.~\cite{HaSh:inprep}, the
second line can be written
\begin{equation}
\frac{\mathcal I f(0)}{4 \pi^2 L} +\mathcal O(e^{-mL})
\,,
\end{equation}
where ${\cal I}$ is a geometrical factor, equal to $Z_{00}(1,0)$
of Ref.~\cite{Luscher:1986n2}.
Combining these results we obtain
\begin{equation}
S_I = \int_{\vec p}^\Lambda \frac{f(\vec p^2)}{\vec p^2}
+ 
\frac{\mathcal I f(0)}{4 \pi^2 L} 
- \frac{f'(0)}{L^3} +\mathcal O(e^{-mL}) \,.
\label{eq:SIres}
\end{equation}
In other words, the sum can be replaced by an integral with 
exponential accuracy aside from a residue consisting of a $1/L$ and a
$1/L^3$ term.

The second sum we need is the generic ``$1/(\vec p^{\,2})^2$" sum
\begin{equation}
S_J = \frac1{L^3} \sum_{\vec p\ne 0}^\Lambda \frac{g(\vec p^2)}{(\vec p^2)^2}
\,.
\end{equation}
This sum can be rewritten as
\begin{align}
S_J &= \frac1{L^3}\sum_{\vec p\ne 0}^\Lambda \frac{g(\vec p^2)-g(0)}{(\vec p^2)^2}
+ \frac{g(0)}{L^3} \sum_{\vec p\ne 0} \frac1{(\vec p^2)^2}\,.
\end{align}
The first term now diverges only as $1/\vec p^{\,2}$ in the IR, so that,
using the result (\ref{eq:SIres}), the sum can be replaced by an integral
up to ${\cal O}(1/L)$ corrections.
The second term is proportional to a sum over integer vectors
${\cal J} = \sum_{\vec n\ne 0} (\vec n^2)^{-2}$
which equals $Z_{00}(2,0)$ in the notation of Ref.~\cite{Luscher:1986n2}.
Thus we obtain
\begin{equation}
S_J = \int_{\vec p}^\Lambda \frac{g(\vec p^2)-g(0)}{(\vec p^2)^2}
+ \frac{L g(0) {\cal J}}{16 \pi^4} + {\cal O}(1/L)
\,.
\label{eq:SJres}
\end{equation}

\section{Scattering length and effective range}
\label{app:effectiverange}

In the main text we need the perturbative expressions for the s-wave scattering length $a$ 
and the effective range $r$ in the scalar $\lambda \phi^4$ theory.
The standard definition of these quantities in terms of the s-wave phase shift $\delta_0$ 
(using the nuclear physics sign convention for $a$) is
\begin{equation}
\frac{\tan{\delta_0(q)}}{q}
= - a \left( 1 + \frac{r a q^2}2 + {\cal O}(q^4)\right)
\,.
\label{eq:B1}
\end{equation}
Here $q$ is the magnitude of the spatial momentum of each particle in the CM (center of mass) frame,
which is related to the total CM energy by $E_2 = 2 \sqrt{q^2+m^2}$.
The phase shift itself is related to the two-particle s-wave scattering K-matrix
(defined as the angular average of the full K-matrix) by
\begin{equation}
{\cal K}_{2,s}(q) = {16 \pi E_2} \frac{ \tan \delta_0}{q}
\,.
\label{eq:B2}
\end{equation}
The K-matrix is related to the usual scattering amplitude by
\begin{equation}
\frac1{{\cal M}_{2,s}(q)} = \frac1{{\cal K}_{2,s}(q)} -i \frac{q}{16\pi E_2}
\,.
\label{eq:B3}
\end{equation}
It follows that the scattering length is given by
\begin{equation}
32 \pi m a = - \mathcal K_{2,s}(0) = - \mathcal M_{2,s}(0)
\,.
\label{eq:B4}
\end{equation}

Through one-loop order, evaluating the diagrams of Fig.~\ref{fig:C2}(c), (d), (e) and
the u-channel version of (e) [treating these as Feynman diagrams for infinite-volume
scattering] one finds
\begin{align}
{\cal M}_{2,s}(0) & = 
-\lambda_0 \left\{
1 - {\lambda_0} A_2 
+ {\cal O}(\lambda_0^3) \right\}
\,,
\label{eq:B5}
\\
A_2 &= \frac12
\int_{\vec p}^\Lambda \left[ \frac1{4 \omega_p \vec p^2} + \frac1{2 \omega_p^3} \right]
\label{eq:A2def}
\,.
\end{align}
Here we have done the integral over the loop energy variable $p_0$ so as to write
the result in the same form as those we obtain in the main text.
The first term in square braces arises from the s-channel loop, while the
second is the sum of those from the t- and u-channel loops.
As usual, the superscript $\Lambda$ indicates an (unspecified) UV regulator.

We find it convenient to adopt a physical renormalization condition for the coupling constant,
defining it in terms of the scattering length as
\begin{align}
\lambda &\equiv -{\cal M}_{2,s}(0) = 32 \pi m a 
\label{eq:lambdadef}
\\
&= \lambda_0
\left(1- {\lambda_0} A_2  +\lambda_0^2 A_3 + {\cal O}(\lambda_0^3) \right)
\label{eq:lambdadef2}
\,,
\end{align}
where $A_3$ is the third-order coefficient that we do not need explicitly.

To obtain the effective range we expand the
K-matrix away from threshold in powers of $q^2$.
Specifically, using Eqs.~(\ref{eq:B1}), (\ref{eq:B2}) and
(\ref{eq:lambdadef2}) we find
\begin{equation}
m^2 \frac{d \mathcal K_{2,s}}{d q^2}\Bigg|_{q^2=0} = -\lambda
\left[
\frac12 + \lambda \frac{rm}{64\pi}\right]
\,,
\label{eq:B8}
\end{equation}
where the first term in the square braces is of kinematic origin,
coming from the expansion of $E_2$ in Eq.~(\ref{eq:B2}).

Dependence on $q^2$ first arises at one-loop order.
For the sum of t- and u-channel loop-diagrams, a straightforward calculation
finds that the scattering is pure s-wave, and yields the contribution
\begin{equation}
m^2 \frac{d \mathcal K_{2,s}}{d q^2}\Bigg|_{q^2=0} 
\supset
- \frac{\lambda_0^2}{48 \pi^2}
\,.
\label{eq:B9}
\end{equation}
Note that the $\mathcal K_2=\mathcal M_2$ for these diagrams,
since there are no physical cuts.

The s-channel loop diagram does have a cut, so $\mathcal K_2$ and $\mathcal M_2$
differ. We elect to calculate the former, which
requires using the principle value prescription for the pole that 
remains after doing the energy-component integral. 
The contribution is purely s-wave, and gives
\begin{equation}
\mathcal K_2 (\text{s-channel})
=  \frac{\lambda_0^2}8 
\text{PV} \int_{\vec p} \frac1{\omega_p(\vec p^2-q^2)}
\,.
\end{equation}
Expanding in powers of $q^2$ we obtain\footnote{%
One must evaluate the principle value integral for $q^2>0$ to
obtain this result. To obtain the correct result with $q^2<0$,
one must use a modified PV prescription that yields analytic
dependence on $q^2$~\cite{Hansen:2014eka}.}
\begin{equation}
\text{PV} \int_{\vec p} \frac1{\omega_p(\vec p^2-q^2)}
=
\int_{\vec p} \frac1{\omega_p\vec p^2}
- \frac{q^2}{2\pi^2} + \mathcal O(q^4)
\,.
\end{equation}
The first term reproduces the s-channel contribution to $A_2$
[cf. Eqs.~(\ref{eq:B5}) and (\ref{eq:A2def})].
The second term leads to the following contribution to the derivative:
\begin{equation}
m^2 \frac{d \mathcal K_{2,s}}{d q^2}\Bigg|_{q^2=0} 
\supset
- \frac{\lambda_0^2}{16 \pi^2}
\,.
\label{eq:B12}
\end{equation}

Combining Eqs.~(\ref{eq:B9}) and (\ref{eq:B12}) we find
\begin{equation}
m^2 \frac{d \mathcal K_{2,s}}{d q^2}\Bigg|_{q^2=0} 
=
- \frac{\lambda_0^2}{12 \pi^2} + {\cal O}(\lambda_0^3)
\,.
\label{eq:B13}
\end{equation}
Comparing to the definition of $r$ in Eq.~(\ref{eq:B8})
we obtain our final result
\begin{equation}
rm = - \frac{32\pi}{\lambda} + \frac{16}{3\pi} + {\cal O}(\lambda)
\,.
\label{eq:effectiverange}
\end{equation}
We stress again that the strange looking $1/\lambda$ term is of
purely kinematic origin. The non-trivial result of the one-loop
calculation is the constant term. 

\section{$1/L^6$ term from L\"uscher's quantization condition}
\label{app:Luscher}

L\"uscher's original work on the two-particle threshold energy shift
presented explicit results only up to ${\cal O}(1/L^5)$~\cite{Luscher:1986n2}.
To compare to our perturbative result we need also the $1/L^6$ term
from the general quantization condition of Ref.~\cite{Luscher:1991n1}.
To our knowledge, the explicit result for this term has not 
been presented elsewhere, so we determine it here.

We start from the quantization condition 
in the form given in Ref.~\cite{Hansen:2014eka}, itself adapted
from the form derived in Ref.~\cite{Kim:2005}.
For vanishing total momentum, the condition is 
[Eq.~(96) of Ref.~\cite{Hansen:2014eka} with $\vec k=0$]:
\begin{equation}
\det\left({\cal K}_2^{-1} + F\right) = 0
\,.
\end{equation}
Here ${\cal K}_2$ and the kinematical function $F$ 
are matrices in angular-momentum space.
${\cal K}_2$ is diagonal, whereas $F$ contains off-diagonal elements.
Close to threshold, s-wave scattering dominates, and one can show that
it is sufficient to truncate the quantization condition to the s-wave alone,
up to ${\cal O}(1/L^{13})$ in the energy shift, 
at which point the $\ell =4$ amplitude is needed.
Thus, for our purposes, the condition reduces to
\begin{equation}
{\cal K}_{2,s}^{-1}  + F_s = 0
\,,
\label{eq:QC2s}
\end{equation}
with
\begin{equation}
F_s = \bigg[\frac1{L^3} \sum_{\vec p} - \; \widetilde{\text{PV}}\!\int_{\vec p}\bigg]
\frac{H(\vec p)^2}{8 \omega_p^2(E_2-2\omega_p)}
\,.
\end{equation}
Here $H$ is the cutoff function discussed in Appendix~\ref{app:sums}, 
$E_2$ is the two-particle CM energy, %(as in Appendix~\ref{app:effectiverange}),
and the $\widetilde{\rm PV}$ pole prescription is a generalized
principle-value prescription introduced in Ref.~\cite{Hansen:2014eka}.
We stress that this condition is identical to the truncated L\"uscher quantization
condition of Ref.~\cite{Luscher:1991n1} up to exponentially suppressed terms.

In our companion analysis of the three-particle quantization condition near threshold,
we derive the threshold expansion of $F_s$ (called $2m \tilde F_{00}$ in that work)~\cite{HaSh:inprep}:
\begin{align}
F_s &= \frac{1}{4 E_2}
\Big\{ \frac1{q^2 L^3} - \frac{\cal I}{4\pi^2 L} - \frac{q^2 L^3 \cal J}{(4\pi^2 L)^2 }
\nonumber\\
&\qquad \qquad- \frac{(q^2 L^3)^2 \cal K}{(4\pi^2 L)^3} + {\cal O}(L^{-3})\Big\}
\,,
\label{eq:Fsexp}
\end{align}
where $q$ is defined as in Appendix~\ref{app:effectiverange}, the 
geometric quantities ${\cal I}$ and ${\cal J}$
are described in Appendix~\ref{app:sums}, and ${\cal K}=Z_{00}(3,0)$ is a third such
quantity (evaluated in Refs.~\cite{Luscher:1986n2,Beane2007}).
The other result we need to apply the quantization condition is the
expansion of $1/\mathcal K_{2,s}$, which, from Eqs.~(\ref{eq:B1}) and (\ref{eq:B2}), 
is
\begin{equation}
\mathcal K_{2,s}^{-1} = - \frac1{16\pi a E_2} \left[1 - \frac{ra}2 q^2 + {\cal O}(q^4) \right]
\,.
\end{equation}

The kinematic relation we need is 
\begin{align}
q^2 &= m \Delta E_2 + \Delta E_2^2/4
\,,
\label{eq:qtoDeltaE}
\end{align}
where $\Delta E_2 = E_2 - 2m$.
We are interested in the solution to Eq.~(\ref{eq:QC2s}) near threshold,
so we expand $\Delta E$ as
\begin{equation}
\Delta E_2 =  \sum_{n=3}^\infty \frac{a_2^{(n)}}{L^n} \,,
\end{equation}
truncating here at the $a_2^{(6)}$ term.
It follows from Eq.~(\ref{eq:qtoDeltaE}) that $q^2$ has a similar expansion
\begin{equation}
q^2 = \sum_{n=3}^\infty \frac{b^{(n)}}{L^n}\,.
\label{eq:q2exp}
\end{equation}

Inserting the expansions of $F_s$ and $\mathcal K_{2,s}^{-1}$ into
the quantization condition (\ref{eq:QC2s}), and using
Eq.~(\ref{eq:q2exp}), we find
\begin{align}
b^{(3)} &= 4 \pi a
\,,\\
\frac{b^{(4)}}{b^{(3)}} &=  -\frac{a}\pi \mathcal I
\,\\
\frac{b^{(5)}}{b^{(3)}} &= \left(\frac{a}{\pi}\right)^2 (\mathcal I^2-\mathcal J)
\,\\
\frac{b^{(6)}}{b^{(3)}} &= 2\pi r a^2 + 
\left(\frac{a}{\pi}\right)^3 (-\mathcal I^3+3\mathcal I\mathcal J-\mathcal K)
\,.
\end{align}
Converting this to a result for $\Delta E$ using Eq.~(\ref{eq:qtoDeltaE}) we
obtain the desired results
\begin{align}
a_2^{(n)} &= \frac{b^{(n)}}m \qquad (n=3-5)
\,,\\
\frac{a_2^{(6)}}{a_2^{(3)}} &= \frac{b^{(6)}}{b^{(3)}}- \frac{\pi a}{m^2}
\,.
\end{align}
The last term in $a_2^{(6)}$ arises from the expansion of the relativistic form of
the energy, i.e. from the second term in Eq.~(\ref{eq:qtoDeltaE}).
\begin{comment}
It is noteworthy that the results for $a_n$ agree with those of Ref.~\cite{Beane2007}
aside from this term.
\end{comment}
The results for $a_n$ agree with those of Ref.~\cite{Beane2007}
aside from this term, suggesting that the source of the discrepancy
is the use of a nonrelativistic dispersion relation in that work.

In the main text we work only to cubic order in $\lambda$. 
At this order we have
\begin{equation}
a_2^{(6)} = \frac{4\pi a}{m} \left( 2 \pi r a^2 - \frac{\pi a}{m^2}  \right)+ {\cal O}(\lambda^4)
\,,
\end{equation}
where $a$ and $r a^2$ both start at $\mathcal O(\lambda)$.

%\newpage
%\bibliographystyle{apsrev} %%% physical review
\bibliography{ref} %%% ref.bib file

\end{document}